\documentclass[preprint,1p,times]{elsarticle}

\usepackage{amsmath}
\usepackage{amssymb}

\usepackage{xcolor}
\definecolor{nozzle_green}{HTML}{008066}
\definecolor{nozzle_yellow}{HTML}{FFFF66}

\usepackage[pdfencoding=auto]{hyperref}
\usepackage{cleveref}
\usepackage[english]{babel}
\usepackage{siunitx}
\usepackage{booktabs}
\usepackage{geometry}
\usepackage{tabulary}
\usepackage{subcaption}
\usepackage{lipsum}
\usepackage[export]{adjustbox}
\usepackage[]{pgfplots}
\usepackage[]{tikz}
\usetikzlibrary {arrows.meta}
\pgfplotsset{compat=1.18}

\DeclareSIUnit\bar{bar}

\newcommand{\mycirc}[1][black]{\textcolor{#1}{\ensuremath\bullet}}

\begin{document}

\begin{frontmatter}

\title{Quantifying water-driven geometric uncertainties in powder bed concrete printing using high-resolution 3D modeling}

\author[inst1]{\texorpdfstring{Christoph Wolf\corref{cor1}}{Christoph Wolf}}
\ead{christoph.wolf@bam.de}
\cortext[cor1]{Corresponding author}

\affiliation[inst1]{organization={Bundesanstalt für Materialforschung und -prüfung (BAM)},
            addressline={Unter den Eichen 87},
            city={12205 Berlin}, 
            state={Germany}}

\author[inst1]{Petr Hlaváček}
\author[inst1]{Annika Robens-Radermacher}
\author[inst1]{Daniel Kadoke}
\author[inst1]{Jörg F. Unger}

\journal{arXiv (cs.CE)}

\begin{abstract}
Dimensional accuracy in powder bed 3D printing of concrete is strongly influenced by binder distribution, and the resulting geometric deviations can be direction-dependent. This study examines how voxel-wise water dosage influences geometric fidelity and deviation anisotropy. Experiments show that small changes in water content can cause large, systematic deviations, including edge rounding and swelling.

We quantify these effects using high-resolution stereophotogrammetry, aligning as-built scans with CAD models. We then compute deviation metrics such as point-wise distance errors and volumetric differences across multiple water-dosage settings, revealing repeatable, directionally biased deformation patterns that intensify with higher water content.

Mechanical testing indicates that stiffness and strength change only marginally, with no clear trend in the tested range. This is explained by excess voxel water diffusing into surrounding powder, leaving the effective water-cement ratio largely unchanged.

Finally, we demonstrate a design-compensation concept that pre-adjusts digital geometry to counter predictable deviations, improving accuracy without post-processing.
\end{abstract}

\begin{graphicalabstract}
{
\setlength{\fboxsep}{0pt}%
\setlength{\fboxrule}{1pt}%
\fbox{\includegraphics[width=0.99\textwidth]{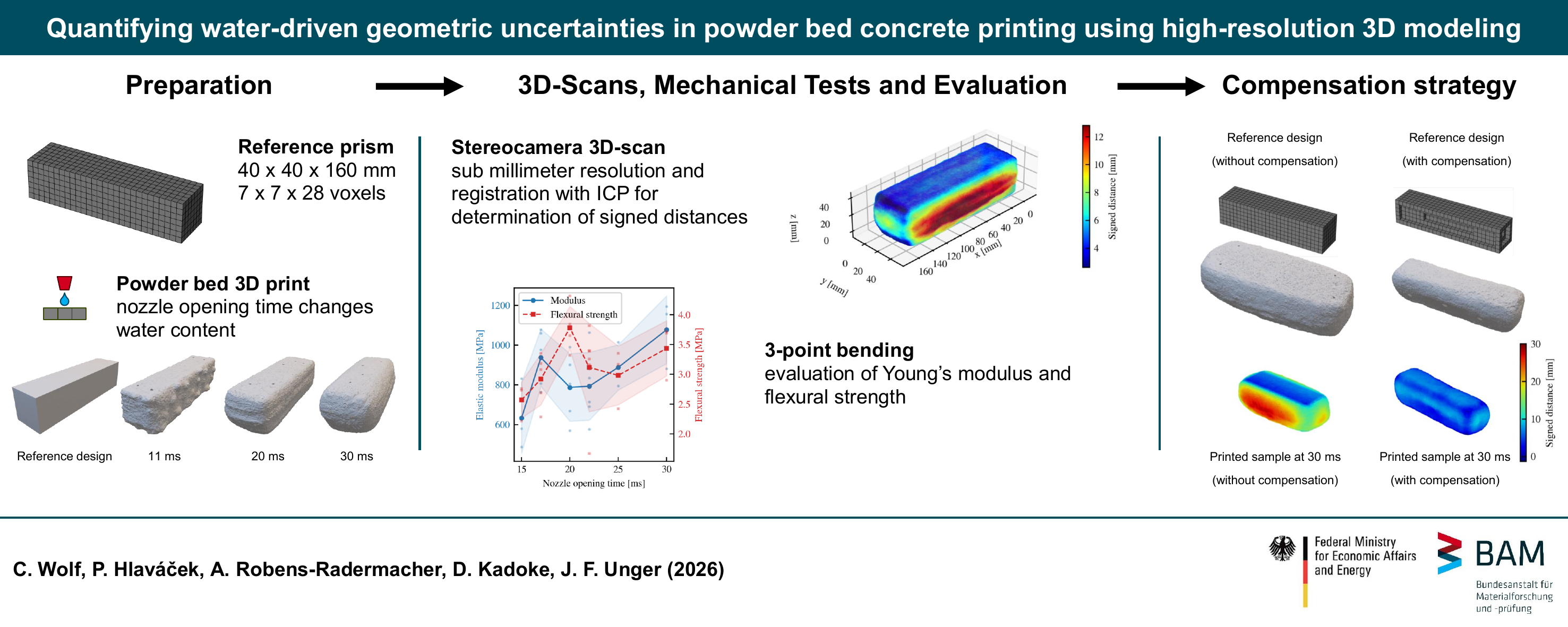}}
}
\end{graphicalabstract}

\begin{highlights}
\item High-resolution 3D modeling enables surface-by-surface deviation analysis.
\item Proposed set of deviation metrics allows systematic geometry deviation assessment.
\item Directional dependency of deviations and water dosage.
\item Mechanical properties stable since extra water spreads without raising w/c ratio.
\item CAD pre-deformation compensates geometric deviations at the design stage.
\end{highlights}


\end{frontmatter}

\section{Introduction}
\label{sec:1_introduction}

Additive manufacturing (AM) technologies have emerged as transformative approaches in the construction industry, offering unprecedented design freedom, reduced material waste, and the potential for automated fabrication of complex geometries \cite{DeSchutter2018_ARTICLE, Buswell2007_ARTICLE}. Among the various AM techniques adapted for cementitious materials, concrete 3D printing (3DCP) has gained substantial attention over the past decade, driven by the need for more sustainable, efficient, and customizable construction methods \cite{Lim2012_ARTICLE,Buswell2007_ARTICLE}. The construction sector accounts for approximately \qtyrange{36}{40}{\percent} of global energy consumption and greenhouse gas emissions, underscoring the urgent need for sustainable construction technologies \cite{Zhuang2024_ARTICLE}. 3DCP technology offers potential solutions through optimized material usage, reduced labor requirements, and the elimination of formwork in many applications \cite{DeSchutter2018_ARTICLE}.

The technology encompasses several distinct process categories, each with unique advantages and limitations depending on application requirements, scale, and material properties. Extrusion-based concrete printing, the most widely adopted approach, deposits material layer-by-layer through a nozzle in a continuous bead \cite{Buswell2018_ARTICLE,Mohan2021_ARTICLE}. This method has demonstrated success in creating structural elements and even full-scale building components, benefiting from relatively straightforward equipment requirements and compatibility with various concrete formulations \cite{Khoshnevis2001_ARTICLE,Khoshnevis2004_ARTICLE,Weger2021_ARTICLE}. However, extrusion-based processes face inherent constraints related to overhang angles, the need for support structures in some geometries, and limitations in achieving fine geometric details \cite{Shakor2019_ARTICLE,Le2012_ARTICLE}. Additionally, the fresh concrete must possess specific rheological properties, sufficient flowability for extrusion while maintaining adequate yield stress and plastic viscosity to support subsequent layers without deformation \cite{Roussel2020_ARTICLE,Nair2020_ARTICLE,Ketel2019_ARTICLE}.

Powder bed 3D printing for concrete, also known as binder jetting or selective cement activation, operates on fundamentally different principles that offer distinct geometric and material advantages \cite{Lowke2018_ARTICLE,Lowke2020_ARTICLE, Diener2021_ARTICLE}. In this process, a thin layer of dry cementitious powder is spread across a build platform, followed by selective deposition of a liquid binder (typically water or water-based solutions) in the desired cross-sectional pattern \cite{Shakor2022_ARTICLE}. This cycle repeats layer-by-layer until the complete component geometry is formed. The process can be further classified into three main techniques: selective binder activation (water deposited on cement and aggregate mixture), selective paste intrusion (cement paste applied to aggregate layers), and binder jetting (binder applied to powder bed with separate activator) \cite{Lowke2018_ARTICLE,Shakor2022_ARTICLE}. The surrounding unbound powder provides inherent support for overhanging features and complex internal geometries, eliminating the need for additional support structures and enabling the fabrication of designs that would be challenging or impossible with extrusion-based methods \cite{Lowke2018_ARTICLE}. Furthermore, powder bed printing allows for precise control over local material properties through variable binder saturation, offers superior surface resolution, and can achieve more intricate geometries due to the fine layer thicknesses achievable (typically \qtyrange{1}{5}{\milli\meter} compared to \qtyrange{10}{50}{\milli\meter} for extrusion) \cite{Salari2023_ARTICLE,Lowke2020_ARTICLE}.

Despite these advantages, powder bed 3D printing for concrete introduces specific challenges related to process control and geometric accuracy \cite{Salari2022_ARTICLE,Mai2022_ARTICLE}. While material properties have been extensively studied for extrusion-based 3DCP through comprehensive interlaboratory efforts \cite{RobensRadermacher2025_ARTICLE,Mechtcherine2025_ARTICLE,Bos2025_ARTICLE}, comparable systematic investigations for powder bed printing methods remain limited. The interaction between powder characteristics (particle size distribution, flowability, composition), binder properties, and deposition parameters significantly influences the hydration kinetics, material consolidation, and ultimately the dimensional accuracy of printed components \cite{Brunner2024_INBOOK,Lowke2022_ARTICLE}. Water dosage, in particular, plays a critical role: insufficient fluid penetration results in weak inter-layer bonding and incomplete hydration, while excessive water content leads to powder migration, bleeding, and geometric distortions \cite{Wang2024_ARTICLE,Salari2022_ARTICLE,Shahid2024_2_ARTICLE}. These process-induced variations manifest as deviations between the intended as-designed geometry and the as-built physical component, affecting both dimensional precision and surface quality-factors that are crucial for applications requiring tight tolerances or aesthetic surface finishes \cite{Xu2020_ARTICLE,Buswell2022_ARTICLE,Shakor2020_ARTICLE}. Advanced measurement techniques have been developed, with \cite{Nair2022_ARTICLE} proposing mathematical morphology-based point cloud analysis for geometry assessment (including the definition of a print accuracy index - PAI), combining global visual comparison with layer-wise quantification using topological methods. Another printibility index is defined in \cite{Ketel2019_ARTICLE} using 3D scanning methods. We employ a comparable analytical framework, focusing specifically on the relationship between water content and geometric fidelity.

For extrusion-based printing, understanding material deformation has been addressed through computational approaches, with \cite{Rizzieri2023_ARTICLE} using the Particle Finite Element Method to study the influence of key process and material parameters to the shape of the free surface of the filament. In \cite{Rizzieri2025_MISC} a deep learning framework was introduced for predicting filament geometry. Experimental compensation strategies have also been developed, with \cite{Ashrafi2021_ARTICLE} establishing relationships between process parameters and deformation, and \cite{Ashrafi2022_ARTICLE} developing grammar-based algorithms for generating compensated toolpaths. These developments demonstrate a mature understanding of geometry prediction and shape accuracy control in extrusion-based methods. For powder bed printing methods, particularly selective cement activation (SCA) and binder jetting, research has focused intensively on understanding the complex interactions between material composition, water dosage, and resulting properties. \cite{Lowke2020_ARTICLE} conducted comprehensive investigations demonstrating that material and process parameters, including w/c ratio, particle size distribution, methylcellulose addition, and water application method, significantly affect both strength and dimensional accuracy. Notably, their work revealed that, contrary to conventional concrete, SCA can exhibit increasing strength with increasing water content within certain ranges. Specifically methylcellulose addition and its influence on shape accuracy has been extensively studied in \cite{Tiwari2024_ARTICLE}. The fundamental physics of fluid penetration in powder beds has been systematically investigated by \cite{Mai2022_ARTICLE}, who developed experimental and analytical frameworks for understanding binder intrusion behavior, while also quantifying the shape accuracy with a height deviation and a volume based coefficient. The authors of \cite{Lowke2022_ARTICLE} synthesized understanding of material-process interactions across the complete SCA process chain. The relationship between mix proportions and print quality has been extensively studied. For powder bed printing, optimal mechanical strength and dimensional accuracy have been demonstrated to result from specific combinations of sand/cement ratio (\num{2.0}), water-cement ratio (\num{0.151}), and sand gradation, though over-saturation and coarser aggregates were found to increase geometric distortions \cite{Wang2024_ARTICLE}. Vertical grading of the w/c ratio (core-shell gradation) has been shown to improve dimensional accuracy compared to uniform high w/c ratios while maintaining compressive strength \cite{Herding2024_INBOOK}. For binder jetting, investigations using binary cement mixtures have demonstrated that optimal proportions of ordinary Portland cement and quick-setting cement significantly influence both mechanical performance and dimensional accuracy, with layer thickness reductions enhancing geometric precision by reducing void formation \cite{Shahid2024_ARTICLE}. Cement content reduction strategies have been explored in selective cement activation processes, revealing that dimensional accuracy increases with decreasing cement content at constant w/c ratio (achieved either through increased aggregate content or substitution with quartz flour), though this improvement comes at the cost of reduced compressive strength which is a trade-off between sustainability goals and structural requirements \cite{Herding2023_ARTICLE}. However, despite these advances in understanding material behavior and process parameters, comprehensive frameworks for predicting and controlling print geometry comparable to the computational approaches predicting filament geometries based on fluid simulations \cite{Rizzieri2023_ARTICLE, Rizzieri2025_MISC, Ashrafi2022_ARTICLE} developed for extrusion-based methods remain underdeveloped for powder bed concrete printing.

In defiance of substantial progress, several research gaps remain. Most quality assessment research has focused on extrusion-based methods, with limited systematic investigation of geometric uncertainties specific to powder bed printing. Furthermore, although relationships between individual process parameters and dimensional accuracy have been established, comprehensive frameworks for predicting and compensating for shape deviations in powder bed systems remain underdeveloped. The issue of geometric precision becomes particularly critical when considering practical applications that require the assembly of multiple printed components. In large-scale construction projects, a complete structure must often be composed of numerous smaller powder bed printed elements, frequently topology-optimized to maximize material efficiency. In such assemblies, even minor geometric deviations can prevent proper assembly or lead to premature structural failure. This constraint is inherent to powder bed printing technology: unlike industrial extrusion printers that can fabricate large monolithic structures, powder bed systems are typically limited to smaller build volumes due to the physical constraints of maintaining large powder beds and the labor-intensive excavation process required to remove unbound material. Consequently, the realization of large-scale structures necessitates a modular approach, where dimensional accuracy of individual components becomes paramount for successful integration. The development of systematic compensation approaches that can be applied directly during digital model preparation before printing begins represents an important opportunity to improve geometric accuracy and reduce material waste in powder bed 3D printing of concrete.

This study addresses these challenges by systematically investigating geometric uncertainties in powder bed 3D printing of concrete, with particular focus on the influence of water dosage on dimensional accuracy and on mechanical properties. To characterize shape deviations comprehensively, concrete prisms were fabricated using powder bed printing with varying water dosage parameters. The outer surfaces of printed specimens were captured using high-resolution 3D scanning techniques (as demonstrated for example in \cite{Xu2020_ARTICLE,Buswell2022_ARTICLE,Nair2022_ARTICLE,Glotz2024_ARTICLE,Keller2025_ARTICLE}), generating detailed point cloud data that enables precise geometric analysis. Afterwards, the samples were mechanically tested to determine strength and stiffness. A Python-based computational workflow was developed to perform rigid registration between the scanned 3D models and the reference digital geometry, followed by calculation of signed distance fields to quantify local deviations across the entire specimen surface. This approach allows for spatially-resolved analysis of shape inaccuracies, revealing patterns and magnitudes of geometric deviations as a function of water dosage and other process parameters. Based on the quantified geometric deviations, a compensation approach is proposed that can be applied during the digital model preparation stage. By incorporating predicted shape deviations into the target geometry before printing, this method has the potential to improve dimensional accuracy. The feasibility of this compensation strategy is demonstrated through a proof-of-concept test, showing its potential for minimizing shape inaccuracies in powder bed 3D printed concrete components.

\section{Experimental programm, materials and methods}

\subsection{Powder Bed Printing System}

\begin{figure}[ht!]
	\captionsetup{justification=centering}
    \centering
    \begin{subfigure}[b]{0.38\textwidth}
        \includegraphics[width=1\textwidth]{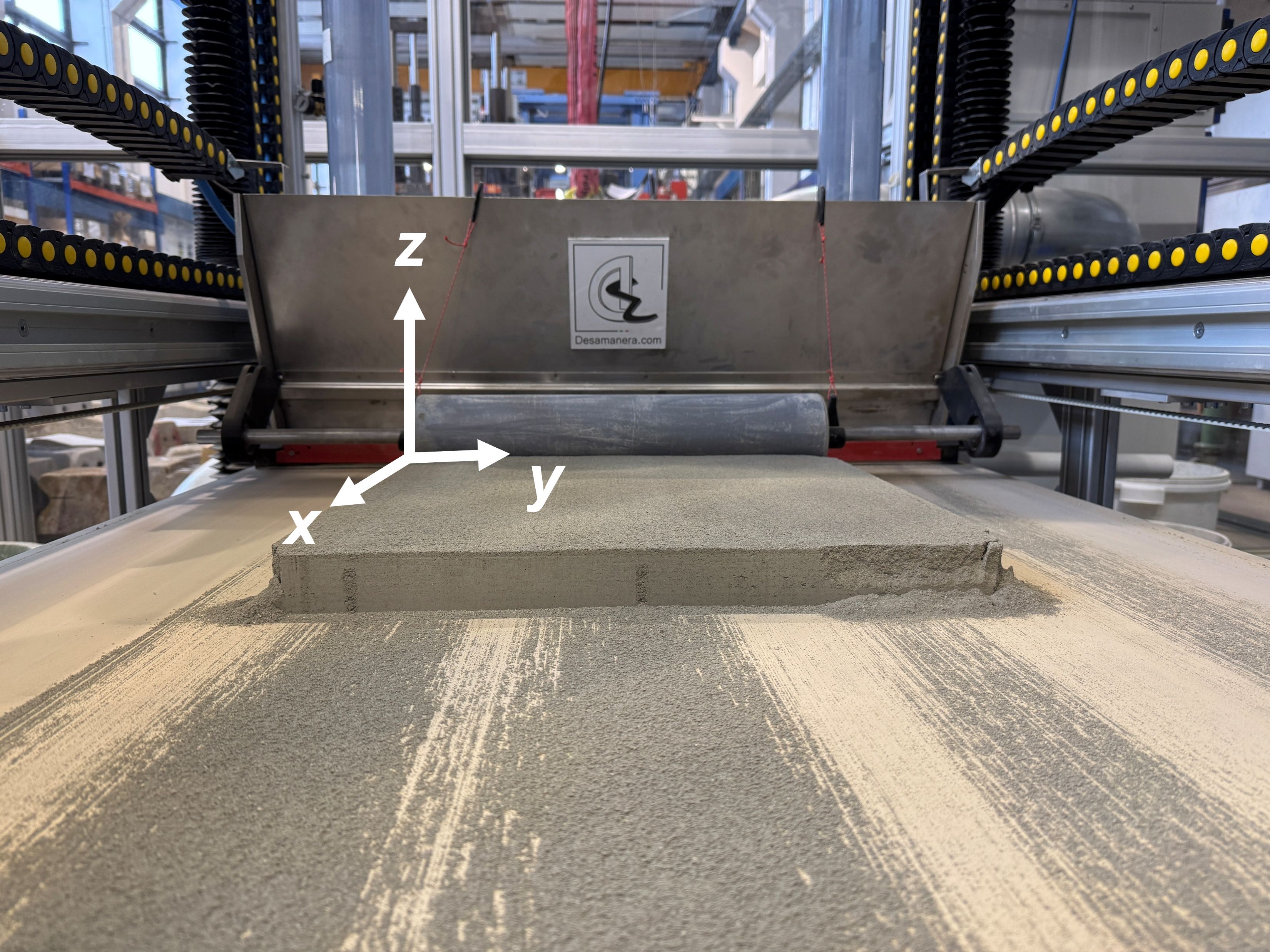}
        \caption{}
        \label{fig:2_printing_results_a}
    \end{subfigure}
	\hfill
    \begin{subfigure}[b]{0.285\textwidth}
        \includegraphics[width=1\textwidth]{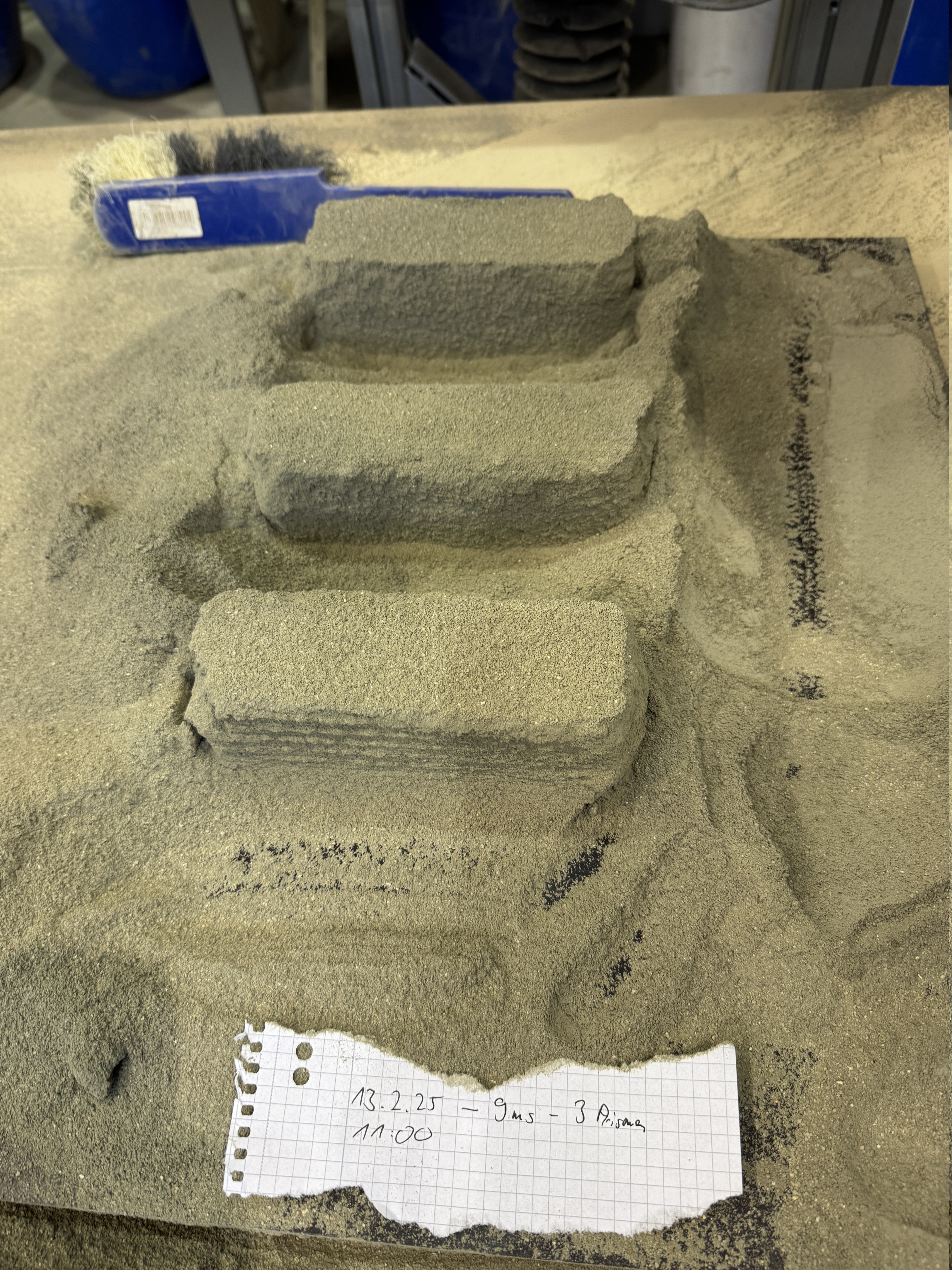}
        \caption{}
        \label{fig:2_printing_results_b}
    \end{subfigure}
	\hfill
    \begin{subfigure}[b]{0.285\textwidth}
        \includegraphics[width=1\textwidth]{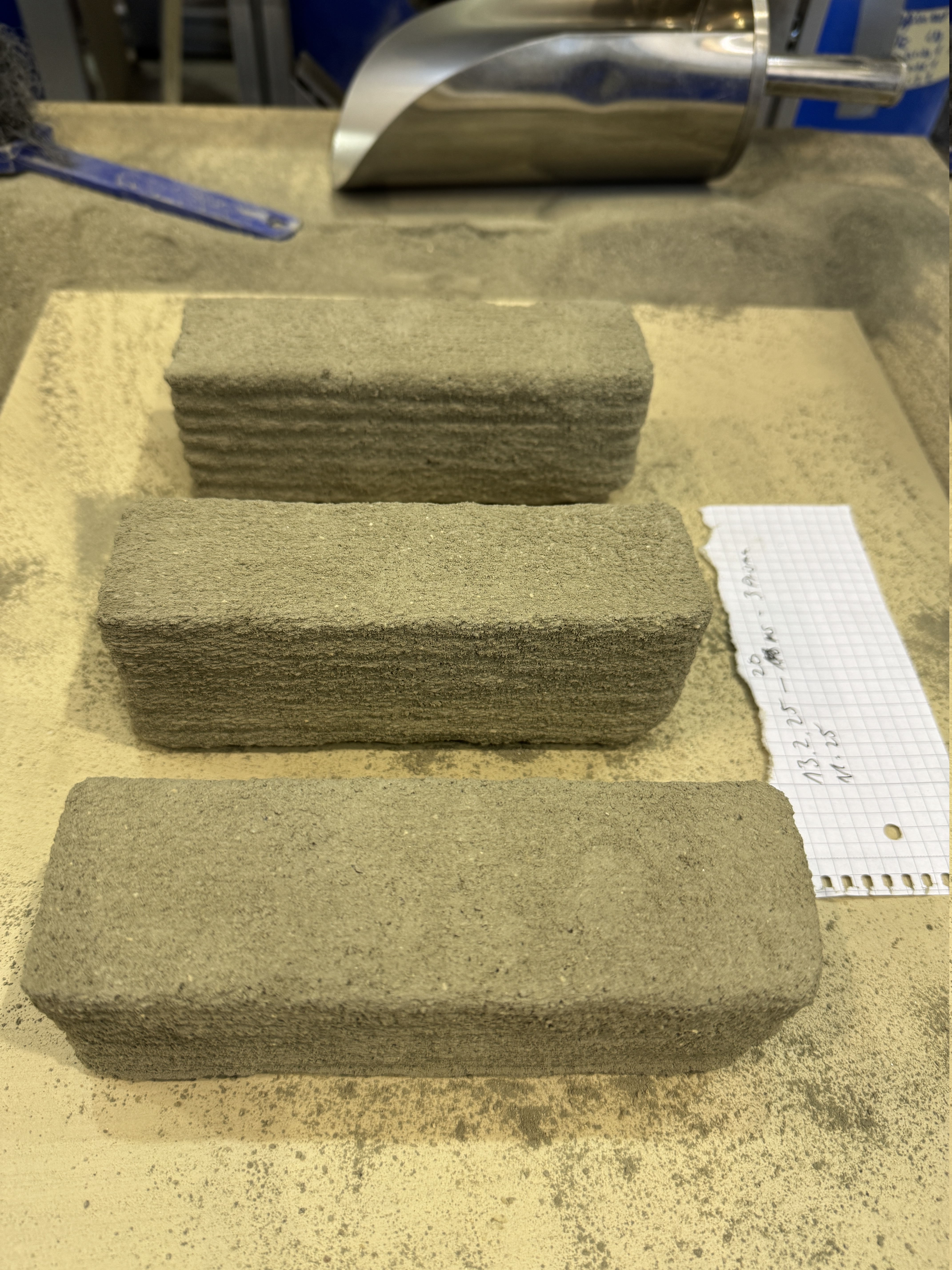}
        \caption{}
        \label{fig:2_printing_results_c}
    \end{subfigure}
    \caption{Typical printing results right after finishing the print. a) The undisturbed powder bed with the prisms enclosed in the powder, not visible from the outside (coordinate system added for reference). b) Excavation process using hands and a soft brush to clean excess powder off the samples. c) Excavated and cleaned prisms ready for additional curing in the autoclave.}
    \label{fig:2_printing_results}
\end{figure}

All specimens were produced using a \textit{DESA1} powder bed 3D printer (manufactured by \textit{Desamanera srl} in Rovigo, Italy). The machine is equipped with a gantry-mounted recoater and nozzle array (see \Cref{fig:2_printing_results_a} and \cite{Salari2024_ARTICLE} for reference). Powder layers were deposited in two passes: in the first pass, half of the target layer thickness was spread in the $+x$-direction, followed by the second pass with a voxel-wise deposition of water and with the deposition of the remaining half of the layer in the --$x$-direction. This sequence allowed freshly deposited dry powder to immediately cover the wetted regions, enhancing water diffusion in the vertical $+z$-direction and promoting a stronger interlayer bond.

Water was supplied hydrostatically from a pressurized reservoir and delivered to the nozzle array through tubing and valves. The printhead contained 184 nozzles arranged in a single row, with an internal nozzle diameter of $\SI{1.1}{\milli\m}$ and a center-to-center spacing of $\SI{5.7}{\milli\m}$. The Droplet volume was controlled by the duration of the electrical opening signal (nozzle opening time), which was adjusted in the printer firmware or in the setup header for each print. The voxel resolution of the system was therefore defined by the nozzle spacing and the chosen layer thickness. For cubic voxels, the layer thickness was set to the nozzle spacing.

The powder mixture used for printing consisted of a cementitious binder and silica sands with different grain size fractions. The binder was Fastcrete\copyright, a CEM I 52.5R cement supplied by Schwenk (Germany), contributing \SI{25}{\percent} by mass of the total mixture ($\alpha_{\mathrm{cem}}=\num{0.25}$). The remaining \SI{75}{\percent} comprised silica sand sourced from the Ottendorf-Okrilla sand depot in Germany, composed of three fractions: \qtyrange[range-units = single, range-phrase = --]{0.1}{0.3}{\milli\m} (\SI{18.75}{\percent}), \qtyrange[range-units = single, range-phrase = --]{0.1}{0.5}{\milli\m} (\SI{18.75}{\percent}), and \qtyrange[range-units = single, range-phrase = --]{0.5}{1.0}{\milli\m} (\SI{37.5}{\percent}). The combination of fine and coarse fractions was selected to optimize the packing density and the layer stability during the deposition process. The resulting powder exhibited a bulk density of approximately \SI{1.7}{\g\per\centi\m\tothe{3}}, ensuring good flowability and uniform layer formation during powder deposition. Water was used as the sole liquid binder, and voxel-wise dosage was varied systematically by adjusting nozzle opening times in the range of \qtyrange[range-units = single, range-phrase = --]{10}{30}{\milli\s}.

After printing, specimens were left in the powder bed for at least \SI{2}{\hour} under ambient laboratory conditions to allow sufficient handling strength to develop. Afterwards, the printed parts were carefully excavated by hand from the surrounding powder using soft brushes and hands to avoid damage. Finally, the samples were subjected to accelerated curing in an autoclave at \SI{186}{\celsius} and \SI{12}{\bar} for \SI{9}{\hour} to promote hydration and strength development.

Typical printing results are shown in \Cref{fig:2_printing_results}, including post-print (\Cref{fig:2_printing_results_a}), excavation (\Cref{fig:2_printing_results_b}), and post-excavation (\Cref{fig:2_printing_results_c}) stages.

\subsection{Test Specimen Design and Water Dosage Levels}

Prismatic specimens with nominal dimensions of \qtyproduct[allow-quantity-breaks]{40 x 40 x 160}{\milli\m} were selected for the experiments in accordance with DIN EN 196-1 \cite{DINEN196-1}. Due to the voxel-based constraints of the printing system (\SI{5.7}{\milli\m} voxel pitch), the printed specimens had final dimensions of approximately \qtyproduct[allow-quantity-breaks]{159.6 x 39.9 x 39.9}{\milli\m} resulting in a total of 28 x 7 x 7 voxels (1372 voxels total per prism). This ensured geometric compatibility with the printer resolution while remaining close to the standardized test geometry. The prisms were oriented with their long axis aligned parallel to the $x$-direction of the build platform (see \Cref{fig:2_printing_results_a} for a reference coordinate system), and three specimens were printed side by side in a single job.

The range of nozzle opening times was determined through preliminary testing. Dosages below \SI{10}{\milli\s} produced highly brittle and porous samples with an insufficient green strength, while values above \SI{35}{\milli\s} resulted in an excessive spreading and a loss of overall geometric definition. Based on these observations, nozzle opening times of 11, 15, 17, 20, 22, 25, and \SI{30}{\milli\s} were selected for a systematic investigation.

At least three samples were printed for each selected nozzle opening time in order to ensure a basic level of reproducibility and to enable a statistical evaluation of geometric and mechanical properties. Some specimens cracked or broke during excavation which reduced the number of usable samples for the subsequent analysis. An overview of the printed and retained specimens is given in \Cref{tab:water_dosages}.

For every nozzle opening time, the amount of water extruded by the printing system had to be determined experimentally, as the water pressure at the inlet showed slight variations. Sixteen nozzles were actuated simultaneously for a total duration of ten seconds and the collected water was weighed with a precision balance. Each measurement was repeated three times, and the average droplet mass was calculated from these triplicates (see \Cref{tab:water_dosages}), allowing us to relate a given nozzle opening time to an effective liquid dosage.

For nozzle times associated with very low water contents not all nozzles produced droplets consistently. This led to larger variations in the actual water content of the printed samples. The increased scatter is visible for the droplet masses and the mass flow rates measured at \SI{11}{\milli\second} and \SI{15}{\milli\second}, which represent the two lowest water dosages. Starting at around \SI{15}{\milli\second} nozzle opening time, a linear relationship between nozzle opening time and droplet mass starts to emerge, indicating a more stable droplet formation regime.

\begin{table}[ht!]
\tymax=400pt
\footnotesize
\caption{Overview of water dosages per printed prism for different nozzle opening times and number of specimens printed and retained for analysis}
	\begin{tabulary}{\linewidth}{@{}LCCCC@{}}
    	\toprule
      	{Nozzle opening time $\left[\si{\milli\s}\right]$} & {Droplet mass $\left[\si{\milli\g}\right]$} & {Mass flow rate $\left[\si[per-mode = symbol]{\milli\g\per\milli\s}\right]$} & {Water mass per prism (28x7x7 voxels) $\left[\si{\g}\right]$} & {Specimens retained $\left[-\right]$} \\
      	\midrule
    	11 & $29.52\pm 0.06$ & $2.684\pm 0.005$ & $40.50\pm 0.08$ & 2 \\
		15 & $29.50\pm 0.12$ & $1.967\pm 0.008$ & $40.47\pm 0.16$ & 2 \\ 
		17 & $33.58\pm 0.17$ & $1.975\pm 0.010$ & $46.07\pm 0.24$ & 4 \\ 
		20 & $40.86\pm 0.03$ & $2.043\pm 0.001$ & $56.06\pm 0.04$ & 3 \\
		22 & $43.30\pm 0.34$ & $1.968\pm 0.016$ & $59.41\pm 0.47$ & 3 \\ 
		25 & $51.33\pm 0.11$ & $2.053\pm 0.004$ & $70.42\pm 0.15$ & 1 \\ 
		30 & $63.94\pm 0.55$ & $2.131\pm 0.018$ & $87.73\pm 0.75$ & 3 \\
		\bottomrule
	\end{tabulary}
	\label{tab:water_dosages}
\end{table}

Since the slicing strategy required for voxel-wise control deviates from conventional binder jetting procedures, a custom Python script was developed to generate printer instructions directly from STL files. This ensured reproducibility across builds and allowed a systematic variation of nozzle opening times while keeping all other parameters constant.

\subsection{Water Cement Ratio Analysis}\label{sec:2_water_cement_ratio_analysis}

The water cement ratio is a central parameter for evaluating the performance of concrete, including concrete produced by powder bed printing. The powder used in this study consists of cement and three sand fractions with different grading curves. Each component contributes twenty five percent by mass and the bulk density of the powder is calculated as $\varrho_{\mathrm{powder}}=\SI[per-mode = symbol]{1695}{\kilogram\per\cubic\metre}$ from the individual component densities. A single water discharge from a nozzle is not necessarily contained in a single voxel but can spread through large parts of the powder bed depending on the applied water amount. To estimate the water cement ratio of the printed structures we applied three complementary methods:
\begin{description}
    \item[Method 1: Theoretical voxel based estimate] This approach assumes that the entire water discharge $m_{\mathrm{w, voxel}}$ remains inside one voxel. The voxel volume $V_{\mathrm{voxel}}$ and voxel mass $m_{\mathrm{voxel}} = \varrho_{\mathrm{powder}} V_{\mathrm{voxel}}$ provide the basis for a purely theoretical water cement ratio:
\begin{equation}\label{eqn:2_method1}
(w/c)_{\mathrm{theo}} = \frac{m_{\mathrm{w, voxel}}}{\alpha_{\mathrm{cem}} \cdot m_{\mathrm{voxel}}}
\end{equation}
where $\alpha_{\mathrm{cem}}$ is the cement mass fraction in the powder. This method ignores any water spreading and therefore provides an upper bound for the ratio.
    \item[Method 2: Mass based estimate after printing and curing] The printed samples were weighed after printing and after autoclave curing. Subtracting the water mass (also derived from the nozzles water discharge) from the total mass $m_{\mathrm{total}}$ yields the solid mass of the sample. Multiplying this mass by the cement fraction $\alpha_{\mathrm{cem}}$ gives an estimate of the cement mass $m_{\mathrm{cem}}$. Dividing the known total water mass of the sample (see \Cref{tab:water_dosages}) by this cement mass gives
    \begin{equation}\label{eqn:2_method2}
(w/c)_{\mathrm{mass}} = \dfrac{m_{\mathrm{w, total}}}{\alpha_{\mathrm{cem}}\left(m_{\mathrm{total}} - m_{\mathrm{w, total}}\right)} = \frac{m_{\mathrm{w, total}}}{m_{\mathrm{cem}}}.
    \end{equation}
    A limitation of this method arises because curing took place in a saturated water atmosphere, which may have introduced additional water uptake.
    \item[Method 3: Geometric volume corrected estimate] The theoretical voxel based ratio from Method 1 can be corrected by accounting for the real printed volume of each prism. Using the meshes from the stereophotogrammetry we obtain the enclosed volume $V_{\mathrm{real}}$ of a prism and compute a volume ratio
    \begin{equation}
\gamma = \frac{V_{\mathrm{real}}}{V_{\mathrm{ref}}}
    \end{equation}
    which reflects how the printed geometry deviates from the ideal prisms volume $V_{\mathrm{ref}}$ in terms of geometry. The corrected water cement ratio becomes
    \begin{equation}\label{eqn:2_method3}
(w/c)_{\mathrm{corr}} = \gamma \cdot (w/c)_{\mathrm{theo}}.
    \end{equation}
    This approach accounts for the fact that water is spread across the entire printed structure rather than remaining confined to the theoretical voxel boundaries.
\end{description}

\subsection{3D Scanning and Geometric Alignment Procedure}

All printed prisms were digitized using an ATOS Triple Scan 2, 5MPx2 system (Zeiss, Germany), which employs a dual-camera configuration and a structured blue-light projector to generate high-fidelity surface models. A measuring volume of \qtyproduct{170 x 140}{\milli\m} was used, enabling the investigation of entire specimens with a maximum point precision of approximately \SI{0.1}{\milli\m}.

During scanning, specimens were placed on a motorized turntable and rotated in \SI{10}{\degree} increments. For each prism, a first sequence of images was captured, after which the specimen was inverted and the process was repeated. To ensure accurate merging of individual scans, adhesive reflector markers were applied to the prism surfaces, compensating for the lack of distinctive geometric features. The scanning software automatically matched the two sequences, and manual post-processing was applied to remove spurious points outside the specimen surface and to improve meshing in regions with incomplete data.

\begin{figure}[ht!]
\centering
\includegraphics[clip, trim = 5cm 3.5cm 5cm 4cm]{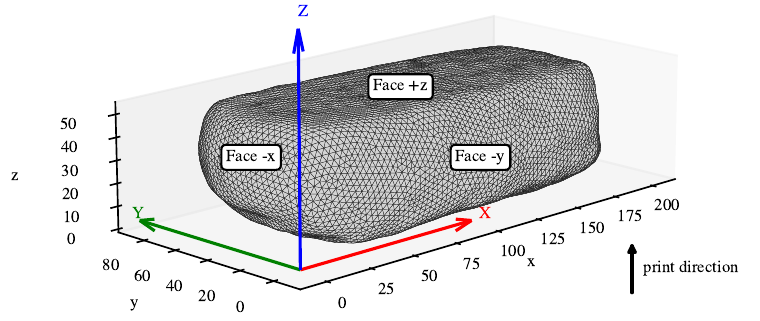}
\caption{Sample 3D scan of a prism with \SI{30}{\milli\s} nozzle time (downscaled to a resolution of \SI{3}{\milli\m} per triangle for illustrative purposes) with coordinate system and face labels for reference}
\label{fig:2_prism_overview}
\end{figure}

The final dataset for each specimen consisted of a watertight surface mesh exported in STL format, accompanied by a vertex cloud for subsequent analysis. The coordinate system was standardized by assigning one corner of the prism as the origin, with the $x$-axis aligned to the long side, the $y$-axis aligned to the short side, and the $z$-axis aligned to the printing direction (see \Cref{fig:2_prism_overview}).

\subsubsection{Alignment to Reference Geometry}\label{sec:2_alignment}

Geometric deviations were quantified relative to a reference prism with dimensions defined by the voxel grid (\qtyproduct[allow-quantity-breaks]{159.6 x 39.9 x 39.9}{\milli\m}). A custom Python workflow was implemented to register the scanned data to this reference geometry.

First, the scanned mesh was downsampled from $\approx$ 2 million to \num{50000} points to balance computational efficiency with accuracy. The reference prism was represented by a surface point cloud generated from \num{1000} random samples, including all eight vertices to stabilize the alignment.

Registration was performed using the Iterative Closest Point (ICP) algorithm as implemented in the \texttt{trimesh}\footnote{\texttt{trimesh} (version 4.8.3). Available at \url{https://github.com/mikedh/trimesh}} package. Given a source point set $P=\left\{p_i\right\}$ and a target point set $Q=\left\{q_i\right\}$, the ICP method iteratively minimizes the mean squared error function $E(R,t)$ between corresponding points by solving:

\begin{equation}
\min \; E(R,t) = \frac{1}{N} \sum_{i=1}^{N} \left\| q_i - \left( \mathbf{R} p_i + \mathbf{t} \right) \right\|^2
\end{equation}

where $\mathbf{R}$ is a rotation matrix and $\mathbf{t}$ a translation vector. Internally the \texttt{trimesh}-package uses a singular value decomposition (SVD) to solve the alignment problem. First, both point sets are centered around their centroid $\bar{p}$ and $\bar{q}$, respectively:

\begin{align}
\tilde{p}_i &= p_i - \bar{p},\\
\tilde{q}_i &= q_i - \bar{q}.
\end{align}

The covariance matrix $\mathbf{H}$ is then computed and the SVD is applied:

\begin{align}
\mathbf{H} &= \sum_{i=1}^{N} \tilde{p}_i \tilde{q}_i^{\mathrm{T}},\\
\mathbf{H} &= \mathbf{U} \mathbf{\Sigma} \mathbf{V}^{\mathrm{T}}.
\end{align}

The optimal rotation $\mathbf{R}$ and translation $\mathbf{t}$ are obtained through iterative computation:

\begin{align}
\mathbf{R} &= \mathbf{V} \mathbf{U}^{\mathrm{T}},\\
\mathbf{t} &= \bar{q} - \mathbf{R} \bar{p}.
\end{align}

\begin{figure}[ht!]
\centering

\begin{subfigure}[b]{0.98\textwidth}
  \includegraphics[trim={0cm, 0cm, 0cm, 0cm}, clip]{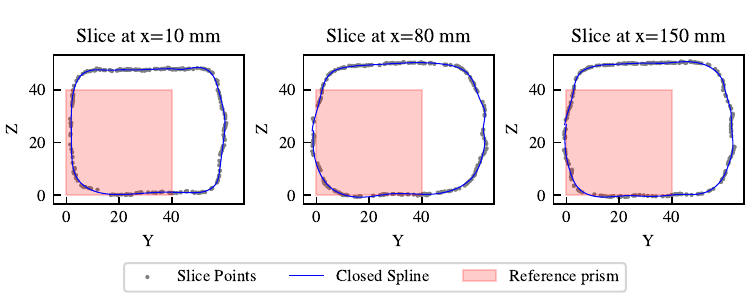}
  \caption{}
  \label{fig:2_prism_not_aligned} 
\end{subfigure}

\medskip
\begin{subfigure}[b]{0.98\textwidth}
  \includegraphics[trim={0cm, 0cm, 0cm, 0cm}, clip]{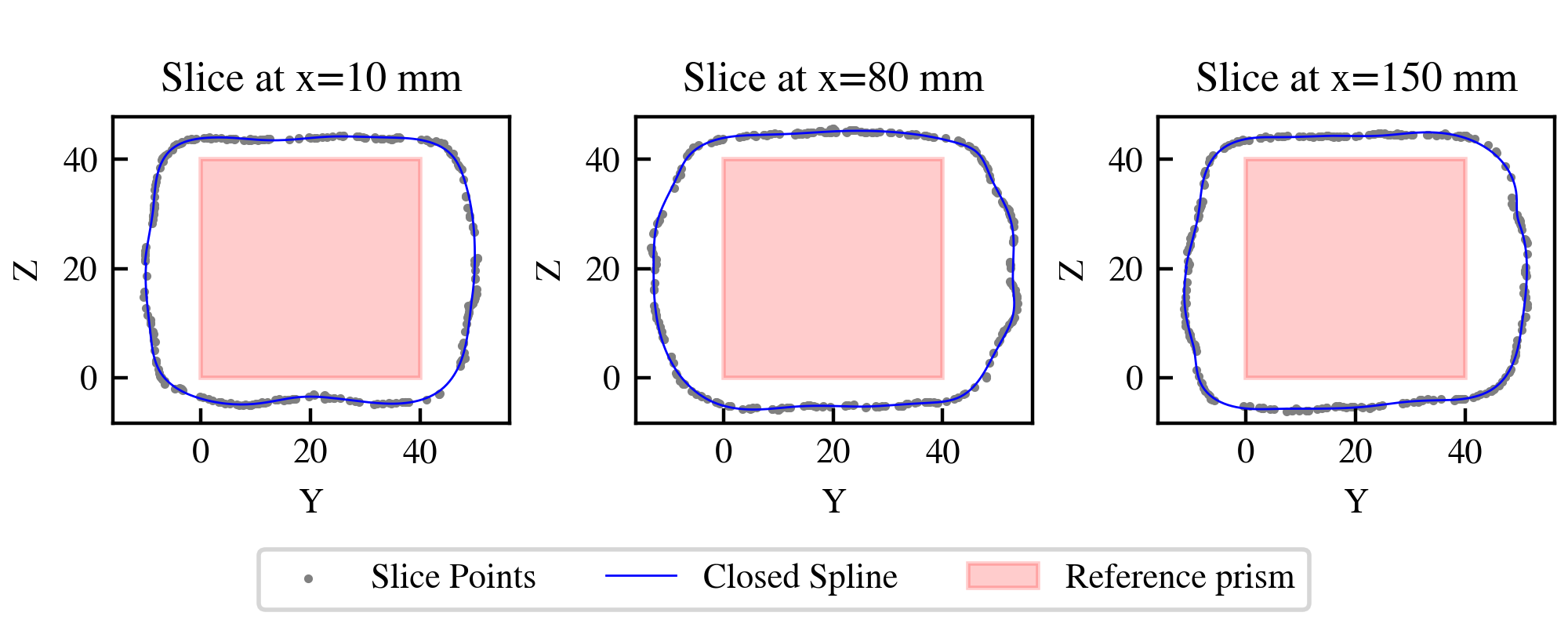}
  \caption{}
  \label{fig:2_prism_aligned}
\end{subfigure}

\caption{Comparison of scanned (blue) and reference (red) geometry before and after registration showing different slices along the $x$-direction of the prism. (a)~The unaligned scanned prism exhibits significant positional and rotational offsets relative to the reference prism. (b)~After applying the ICP registration, the scanned data is closely aligned with the reference, enabling accurate point-wise deviation analysis.}
\label{fig:2_prism_registration}
\end{figure}

A maximum of 100 ICP iterations was set, with convergence determined when either this limit was reached or the change in cost fell below \num{1e-5}. This approach proved robust and more accurate than tested alternatives such as pre-alignment via principal component axes.
A sample registration using this algorithm for a prism with a nozzle time of \SI{30}{\milli\s} is illustrated in \Cref{fig:2_prism_registration}. It is evident that the unaligned scanned geometry (\Cref{fig:2_prism_not_aligned}) exhibits significant positional and rotational offsets relative to the reference prism. After applying the ICP registration (\Cref{fig:2_prism_aligned}), the scanned data is closely aligned with the reference, enabling an accurate point-wise deviation analysis. The alignment process required approximately 40 seconds for this sample with \num{50000} downsampled points on a single core of an AMD EPYC\texttrademark{} 74F3 CPU.

\subsubsection{Calculation of Geometric Deviations and Evaluation Metrics}
\label{sec:2_metrics}

Once registered, the signed distance $d(p)$ of each scanned point $p$ to the reference surface $S$ was computed as:

\begin{equation}
d(p) =
\begin{cases} 
+ \min_{s \in S} \| p - s \|, & \text{if $p$ lies outside the reference surface} \\[6pt]
- \min_{s \in S} \| p - s \|, & \text{if $p$ lies inside the reference surface}
\end{cases}
\end{equation}

This yielded a spatially resolved measure of material excess (positive values) or material loss (negative values). The resulting deviation distributions were exported for further statistical evaluation. The downsampled point cloud was used for this analysis as it showed no significant differences compared to the full-resolution scan.

For a separate evaluation, the deviations were grouped by surface orientation ($x$-, $y$- and $z$-faces of the prism) and analyzed separately. Each scanned point was assigned to its nearest mesh face to determine the corresponding surface orientation. This allowed for detecting anisotropic effects related to printing direction or voxel arrangement.

Several distance-based and statistical metrics were computed from the measured signed distance data to quantitatively assess the geometric accuracy of the printed prisms and their dependence on the nozzle time. These metrics provide compact scalar values that enable a more objective and comparable evaluation of printing quality, rather than relying solely on complex spatial distributions.

Global deviation metrics were employed to quantify the overall geometric agreement between the printed and reference geometries. The Hausdorff distance $d_{\mathrm{H}}$ measures the largest deviation between two point sets, in this case the measured point cloud $P$ and the reference surface $Q$, and is defined as

\begin{equation}
d_{\mathrm{H}}(P, Q) = \max \left\{ \sup_{p \in P} \inf_{q \in Q} \| p - q \|, \; \sup_{q \in Q} \inf_{p \in P} \| q - p \| \right\}.
\end{equation}

This metric captures the worst-case deviation, making it particularly sensitive to local outliers or defects. Complementary to this, the Chamfer distance $d_{\mathrm{C}}$ provides an averaged measure of geometric discrepancy and is calculated as

\begin{equation}
d_{\mathrm{C}}(P, Q) = \frac{1}{|P|} \sum_{p \in P} \min_{q \in Q} \| p - q \| + \frac{1}{|Q|} \sum_{q \in Q} \min_{p \in P} \| q - p \|.
\end{equation}

with $|\bullet|$ being the cardinality (amount of points) of a set. In contrast to the Hausdorff distance, the Chamfer distance reduces the influence of single extreme deviations and thus provides a more robust characterization of the overall shape fidelity.

Additionally, the Print Accuracy Index (PAI),
introduced by \cite{Nair2022_ARTICLE}, was computed to combine local geometric deviations and volumetric fidelity into a single normalized accuracy measure. First, both the scanned point cloud and the reference mesh are aligned with respect to their centroids (after a previous ICP registration with translation and rotation). Then, the distance from the centroid to each point in the scanned cloud ($d_{\mathrm{cs}}$) and the corresponding distance from the centroid to the reference mesh ($D_{\mathrm{cr}}$) are calculated. The ratio of these distances is computed for each point, and these ratios are averaged over all $n$ points of the scanned cloud:

\begin{equation}
    \mathrm{PAI} = \frac{1}{n} \sum_{i=1}^{n} \frac{d_{\mathrm{cs},i}}{D_{\mathrm{cr},i}}.\label{eqn:2_PAI_std_dev}
\end{equation}

Analogous to the calculation of a mean value, the PAI represents the average of the distance ratios across all points. To gain additional insight into the distribution of geometric deviations, the standard deviation of these ratios was also computed:

\begin{equation}
    s_{\mathrm{PAI}} = \sqrt{\frac{1}{n-1} \sum_{i=1}^{n} \left( \frac{d_{\mathrm{cs},i}}{D_{\mathrm{cr},i}} - \mathrm{PAI} \right)^2 }.\label{eqn:2_s_PAI}
\end{equation}

This provides a measure of the variability in the distance ratios and reveals inconsistencies that may be masked by the mean value alone. A scanned point cloud that closely matches the reference mesh would exhibit a PAI near 1.0 and a standard deviation close to zero. However, structures with significant deviations, where points inside and outside the mesh cancel each other out, could still yield a PAI of approximately 1.0, while the standard deviation would reveal the underlying geometric inconsistencies. Together, the PAI and its standard deviation enable a relative comparison of samples produced under different processing conditions, such as varying nozzle times or water dosages. By condensing complex three-dimensional deviation data into these statistical measures, quantitative ranking of process parameter combinations with respect to their geometric accuracy becomes possible.

Overall, the combination of these metrics offers a comprehensive framework for evaluating geometric precision in powder-based concrete 3D printing. The global distances (Hausdorff and Chamfer) reflect overall conformity to the design geometry, while the PAI and its standard deviation allow for more nuanced interpretations of accuracy trends and the identification of systematic effects related to water dosage and voxel control.

\subsubsection{Projection of Signed Distances onto Reference Surfaces}\label{sec:2_projection}

The signed distances between the measured prism surface and the corresponding ideal reference geometry were initially computed for each point in the scanned point cloud, resulting in one signed distance value per measurement point. While this approach provides high spatial resolution, it complicates a systematic evaluation of deviations per individual prism surface ($\pm x$, $\pm y$, $\pm z$). To facilitate a face-wise statistical and spatial analysis, the signed distances were projected onto the respective reference surfaces of the ideal prism geometry.

For this purpose, a regular rectangular sampling grid was generated on each of the six reference surfaces. For every grid node $s_{ij}$ on a given surface $S$, the nearest measured point $p_k$ in the point cloud $P$ was identified according to the minimum Euclidean distance criterion:

\begin{equation}
p_k = \arg \min_{p \in P} \| p - s_{ij} \|.
\end{equation}

The signed distance value $d(p_k)$ of this nearest point was then assigned to the grid node as

\begin{equation}
d_{\mathrm{proj}}(s_{ij}) = d(p_k).
\end{equation}

This procedure effectively transforms the irregularly distributed point cloud data into a structured format aligned with the geometric surfaces of the ideal prism. The resulting gridded data, using identical grid resolutions and coordinates across all specimens, provide a consistent basis for face-specific analysis and enable direct comparison between different process parameters. Mean values and standard deviations of signed distances can be computed for each prism face and water dosage level (nozzle time), and contour plots can be generated to illustrate the spatial distribution of geometric deviations across each surface.
Optionally, the gridded data can be further refined by interpolating the signed distance field between sampling points to obtain a continuous representation of the deviation field.

In practice, this nearest-neighbor projection approach occasionally encountered mismatches that compromised the spatial coherence of the resulting grid. For certain grid nodes, the identified nearest point belonged to a non-adjacent region of the measured point cloud, most commonly to the top or bottom edges of the prism, rather than to the corresponding surface region. This occurred when edge points were marginally closer in Euclidean distance despite not being geometrically relevant to the surface in question. Such mismatches introduced localized artifacts in the projected distance field and occasionally led to physically implausible deviation patterns. To mitigate this issue, a more constrained nearest-neighbor search could be implemented, for instance by restricting the search space to points whose surface normals align with the reference surface normal.

\subsection{Mechanical Testing}
\label{sec:2_mechanical_testing}

The mechanical performance of the printed specimens was evaluated by determining both flexural and compressive strengths following the principles of the European standard DIN EN 196-1 \cite{DINEN196-1}. All tests were performed using a Toni Technik testing machine (Model 1544, Berlin, Germany), equipped with interchangeable fixtures for three-point bending and compression testing. The testing device recorded the applied force and the displacements were measured using an external displacement sensor, which served as the basis for calculating stress and strain. All loads were applied under displacement-controlled conditions at a constant rate of \SI{0.01}{\milli\m/\s}, ensuring a uniform and stable loading process. For the supports in the three-point bending tests, cylindrical rollers were used to minimize friction and allow for slight rotations during loading, thus ensuring a more accurate representation of pure bending conditions.

Flexural strength was determined using three-point bending tests on prism specimens with nominal dimensions of \qtyproduct[allow-quantity-breaks]{40 x 40 x 160}{\milli\m}, in accordance with DIN EN 196-1. The distance between supports was $L_{\mathrm{span}} = \SI{120}{\milli\m}$. The orientation of the prisms during testing replicated their printing configuration, meaning that the applied load acted orthogonal to the printed layers, i.e. parallel to the z-axis. For a geometrically perfect prism this setup would lead to maximum tensile stresses at the bottom face of the specimen, which were used as the basis for further calculations (see \Cref{eqn:2_stress_and_strain}).

For specimens exceeding the nominal prism size, samples were cut to \qtyproduct[allow-quantity-breaks]{40 x 40 x 160}{\milli\m} prior to testing to maintain consistency. For all samples, the actual measured dimensions were used in all subsequent stress and strain calculations.

Based on elastic beam theory, the normal stress $\sigma_x$ and axial strain $\varepsilon_x$ in the three-point bending test were determined from the applied load $F$, the specimen span $L_{\mathrm{span}}$, the specimen width $b$, height $h$, and the mid-span deflection $\delta$ according to:

\begin{equation}\label{eqn:2_stress_and_strain}
\sigma_x = \dfrac{3FL_{\mathrm{span}}}{2bh^2} \quad \text{and} \quad \varepsilon_x = \dfrac{6 h \delta}{L_{\mathrm{span}}^2}.
\end{equation}

These relations assume linear elastic behavior and small deflections, which is valid for the pre-peak regime of all tested prisms.

Compressive strength tests were performed on the two halves of the broken specimens remaining from the flexural strength tests. The tests were conducted between the compression platens of the same Toni Technik system. While the reused halves may contain minor pre-damage from prior loading, comparative tests on separate, untested specimens have demonstrated that the resulting influence on compressive strength is negligible. This approach thus ensures efficient use of printed material without compromising accuracy. For compression testing, the axial stress $\sigma$ and strain $\varepsilon$ were calculated using the measured load $F$, the specimen cross-sectional area $A$, the measured specimen height $h$, and the displacement $\delta$.

\begin{figure}[ht]
    \footnotesize
    \centering
    \begin{tikzpicture}
        \begin{axis}[
            width=0.6\textwidth,
            height=0.5\textwidth,
            xlabel={Strain $\varepsilon_x$ [-]},
            ylabel={Stress $\sigma_x$ [MPa]},
            xmin=0, xmax=0.008,
            ymin=0, ymax=5,
        ]
        \addplot[black, thick] table [x expr=\thisrowno{0}/100, y index=1, skip first n=4, col sep=tab] {bending_20250213_20ms_1_Probe_01.txt};

        \addplot[domain=0.003:0.0055, dashed, red, thick] {796.8*x - 0.95} node[midway, sloped, above] {linear fit (slope $=E$)};

        \coordinate (A) at (axis cs:0.000,2);
        \coordinate (B) at (axis cs:0.0025,2);
        \coordinate (A_L) at (axis cs:0.000,0.1);
        \coordinate (B_L) at (axis cs:0.0025,0.8);
        \coordinate (C) at (axis cs:0.0025,0.3);
        \coordinate (D) at (axis cs:0.006,0.3);
        \coordinate (D_L) at (axis cs:0.006,3.8);

        \draw[Latex-Latex, thick, black, align=center] (A) -- (B) node[midway, above] {preloading\\ phase};
        \draw[thin, black, dashed] (A) -- (A_L);
        \draw[thin, black, dashed] (B) -- (B_L);

        \draw[Latex-Latex, thick, black] (C) -- (D) node[midway, above] {elastic region};
        \draw[thin, black, dashed] (D) -- (D_L);
        \draw[thin, black, dashed] (C) -- (B_L);

        \coordinate (MaxPoint) at (axis cs:0.0066,4.315397936);
        \coordinate (ArrowStart) at (axis cs:0.006,4.315397936);

        \draw[-Latex, thick] (ArrowStart) -- (MaxPoint)
      node[left=0.7cm] {$\sigma_{x,\mathrm{max}}$};

        \end{axis}
    \end{tikzpicture}
    \caption{Overview of the typical stress-strain relation in a 3-point-bending test and illustration of loading regions and key values}
    \label{fig:2_stress_strain_curve}
\end{figure}
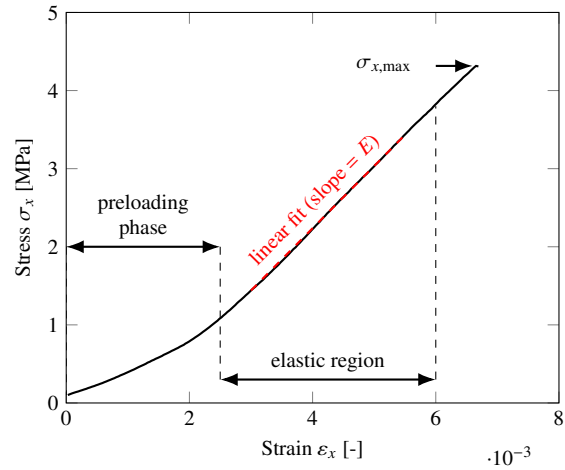

All specimens were loaded until failure, and the peak load was recorded automatically by the control system. The flexural and compressive strengths were computed from the corresponding peak stresses according to the above relations.
Stress-strain curves were derived from the recorded load-displacement data for all tests, enabling detailed analysis of elastic moduli, peak strengths, and failure modes (see \Cref{fig:2_stress_strain_curve} for an overview).

The Young's modulus of the printed specimens was determined from the stress-strain curves obtained during the flexural tests. All samples exhibited similar curve characteristics, with two distinct phases clearly identifiable: an initial preloading phase followed by a predominantly linear elastic phase up to failure.

In the preloading phase, a reduced slope of the stress-strain curve was observed. This initial nonlinearity is attributed to the progressive tightening of internal components within the testing system and the gradual establishment of full contact between the specimen and the loading fixtures. As a result, this region does not represent the intrinsic material behavior and was excluded from the stiffness evaluation, simply by having lower a smaller slope for the linear fit.

Following preloading, the curves exhibited a linear elastic region, extending almost entirely up to the point of failure, corresponding to the maximum flexural stress (tensile strength in bending). Within this linear range, several overlapping stress-strain intervals ('windows') of 100 consecutive data points were defined and fitted with linear regression models. For specimens containing typically 200-400 data points, this window size enabled automated identification of the most linear region, analogous to manual selection. The window yielding the highest coefficient of determination ($R^2$) was selected, and the slope of this regression line was taken as the Young's modulus ($E$) of the specimen (see red dashed line in \Cref{fig:2_stress_strain_curve}).

This approach ensures that the calculated modulus reflects the most representative linear response of the material, minimizing the influence of experimental noise and minor deviations from ideal elasticity.
\newpage

\section{Results and discussions}

\subsection{Overview}

\begin{figure}[ht!]
\centering
\includegraphics[width=0.9\textwidth,clip,trim = 3cm 9cm 3cm 9cm]{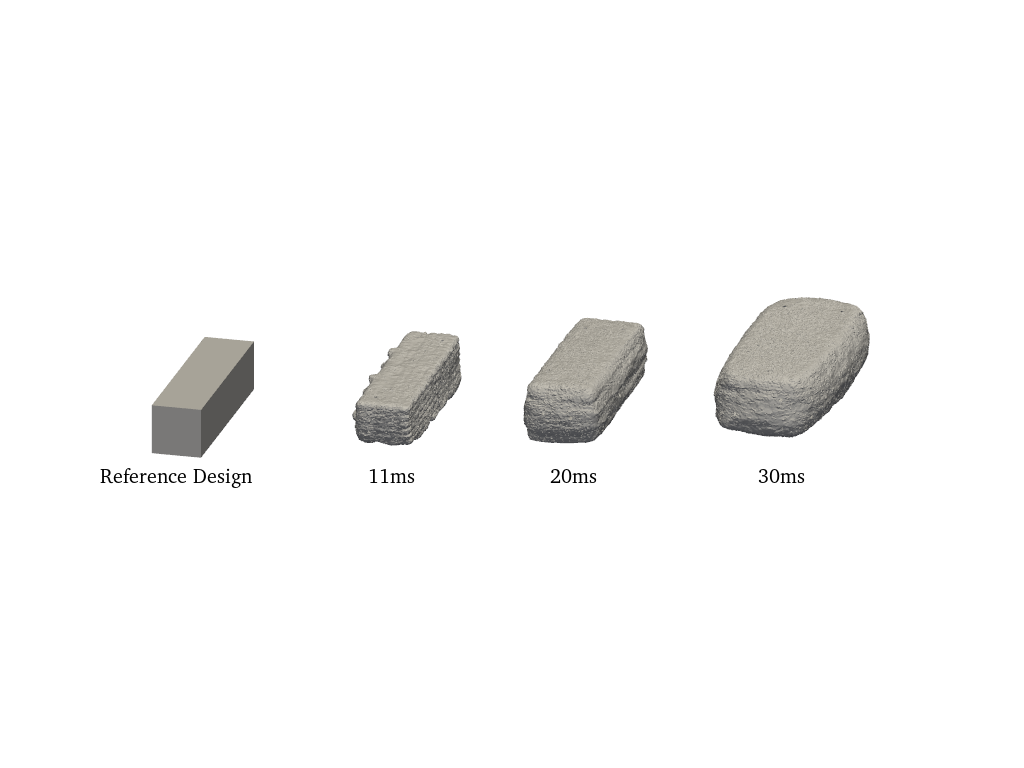}
\caption{Comparison between different water contents, starting with a nozzle opening time of \SI{11}{\milli\s}, to \SI{20}{\milli\s}, up to \SI{30}{\milli\s} (consistent scaling across all prisms)}
\label{fig:results:comparison_different_dosages}
\end{figure}

The effect of water dosage on the printed geometry is qualitatively illustrated in \Cref{fig:results:comparison_different_dosages}, which presents a series of simplified representations comparing the intended CAD shape to specimens printed with increasing water content (\SI{11}{\milli\s}, \SI{20}{\milli\s} and \SI{30}{\milli\s} nozzle opening time respectively). The CAD model features sharp, well-defined rectangular edges, serving as a geometric baseline.

With increasing water dosage, the printed shapes exhibit progressive deformation. At \SI{11}{\milli\s}, the specimen begins to show slight edge softening and minor surface irregularities. At \SI{20}{\milli\s}, rounding becomes more pronounced, and the cross-section starts to deviate from its rectangular profile. Finally, at \SI{30}{\milli\s}, the geometry is significantly distorted: edges are completely rounded, and the cross-section approaches an oval or pill-like shape, indicating extensive material spreading and surface smoothing due to excess water spreading beyond the intended voxels. This trend reflects a clear correlation between water dosage and loss of geometric fidelity. These visual trends are consistent across multiple samples and are further supported by the quantitative deviation metrics presented in the following sections. The observed behavior underscores the critical role of water dosage in determining not only mechanical properties but also the dimensional accuracy of printed parts.

The geometric fidelity of the printed specimens was evaluated by conducting high-resolution 3D scans of the as-printed elements and aligning the resulting point clouds with their corresponding CAD models using an Iterative Closest Point (ICP) algorithm introduced in \Cref{sec:2_alignment}.

\begin{figure}[ht!]
\centering
\includegraphics[clip, trim = 0.3cm 0.7cm 1.4cm 1.6cm]{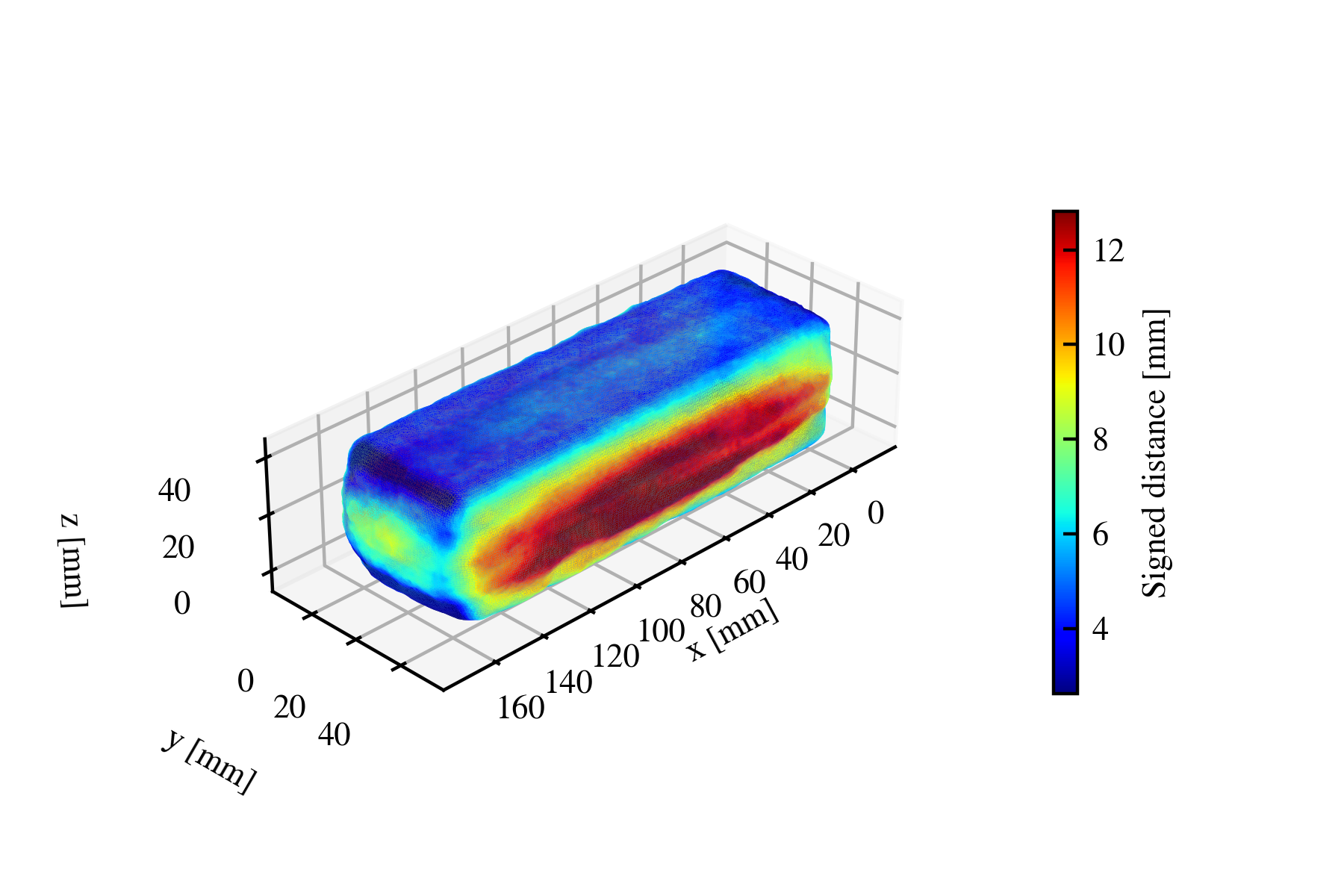}
\caption{Comparison between CAD and as-printed geometry using the \SI{20}{\milli\s} nozzle time}
\label{fig:results:comparison_CAD_3D}
\end{figure}

\Cref{fig:results:comparison_CAD_3D} shows a representative result of this comparison using a color-coded deviation heatmap projected onto the surface of the printed specimen. The color scale represents the signed distance error between the scanned surface and the CAD model, with higher values (red) indicating material excess (i.e., swelling or over-deposition) and lower values (blue) indicating a close-to-reference geometry. The visual representation reveals distinct regions of deviation, with pronounced edge rounding, surface swelling, and localized feature loss, particularly in areas with sharp corners.

\begin{figure}[ht!]
\centering
\includegraphics[]{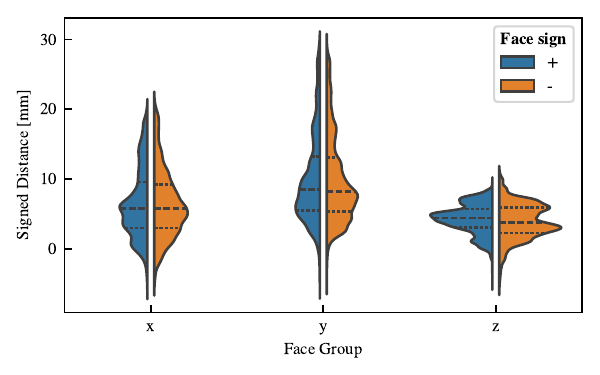}
\caption{Signed point-wise distance distribution between printed and CAD surfaces, grouped by face orientation ($x$, $y,$ $z$). Positive values indicate surface expansion, negative values indicate material loss. Data is aggregated over all samples and split by sign to visualize asymmetry in deformation behavior.}
\label{fig:results:violinplot_faces}
\end{figure}

The six faces of the printed prism were grouped into three categories based on their alignment with the global coordinate system to analyze how geometric deviations vary with surface orientation: $x$-faces ($\pm x$), $y$-faces ($\pm y$), and $z$-faces ($\pm z$), see \Cref{fig:2_prism_overview} for reference. For each group, the signed point-wise distance between the printed surface and the CAD model was calculated, with positive values indicating outward deformation and negative values indicating material loss or surface recession. \Cref{fig:results:violinplot_faces} presents the resulting distributions as a violin plot, split by deviation sign. Each violin shows the full distribution of signed distances across all samples within the respective face group, enabling a clear comparison of deformation tendencies across orientations. Across all tested samples, similar patterns were observed: deviations were generally most significant along the $x$-direction (direction of the nozzle movement) and $y$-direction (parallel to the printer bed), suggesting subsequent spreading of the water through the loose powder. The magnitude of deviation varied systematically with water dosage, as discussed in the subsequent sections.

The results reveal distinct direction-dependent patterns:

\begin{itemize}
\item $x$- and $y$-faces exhibit wide, asymmetric distributions with a clear skew toward positive values (up to $\SI{30}{\milli\m}$), indicating a strong tendency toward outward bulging or material accumulation along the horizontal plane. This behavior is consistent with gravity-driven lateral spreading during the printing process, particularly in response to increased water dosage. Notably, the faces within a single group ($+x$ vs. $-x$ and $+y$ vs. $-y$) show very similar distributions, both qualitatively and quantitatively.
\item $z$-faces, representing top and bottom surfaces, show a narrower and more symmetric distribution centered around smaller deviation values. These faces are less affected by spreading, likely due to vertical constraints during layer deposition and compaction by overlying material and very likely due to the presence of the printer bed for the first layer. 
\end{itemize}

\subsection{Water Cement Ratio Comparison}

\begin{figure}[ht!]
\centering
\includegraphics{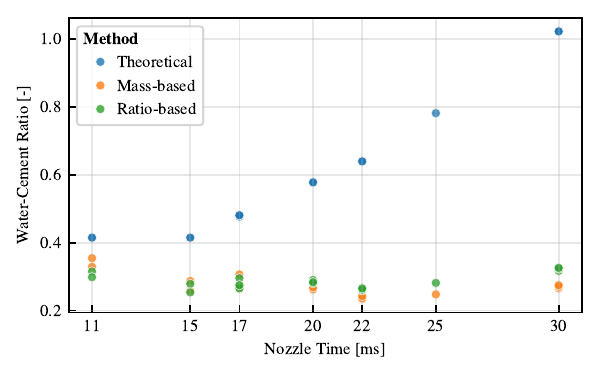}
\caption{Comparison of the theoretical, mass-based and ratio method for calculating the water-to-cement ratio over nozzle times.}
\label{fig:3_water_cement_ratio_comparison}
\end{figure}

Based on the previous calculation procedures and assumptions outlined in \Cref{sec:2_water_cement_ratio_analysis}, \Cref{fig:3_water_cement_ratio_comparison} presents a comparison of the water cement ratios obtained for all available nozzle times. The theoretical water cement ratio (\Cref{eqn:2_method1}) shows a steady increase with increasing nozzle time, which is expected because it relates the added water to a fixed and idealised reference volume with a constant cement mass. While the theoretical values may still be reasonable for low nozzle times, they become increasingly unrealistic for higher water contents and even exceed a ratio of one for the longest nozzle openings.

In contrast, the mass based (\Cref{eqn:2_method2}) and ratio based (\Cref{eqn:2_method3}) methods produce water cement ratios that remain almost constant across the tested nozzle times. All values lie within a range between \num{0.2} and \num{0.4}, with the highest ratios observed for the lowest nozzle times. This behaviour appears more plausible for a cement based system, since powder bed 3D printing relies on similar hydration and hardening mechanisms as conventional concrete mixtures. The comparison between the two practical methods further suggests that the water atmosphere inside the autoclave did not significantly influence the final water cement ratio, because both approaches yield similar values within the same range, assuming that they are comparable.

\subsection{Water Dosage vs. Geometry}\label{sec:3_water_dosage_vs_geometry}

\begin{figure}[ht!]
\centering
\includegraphics[clip, trim=0.3cm 0.3cm 0.29cm 0.5cm]{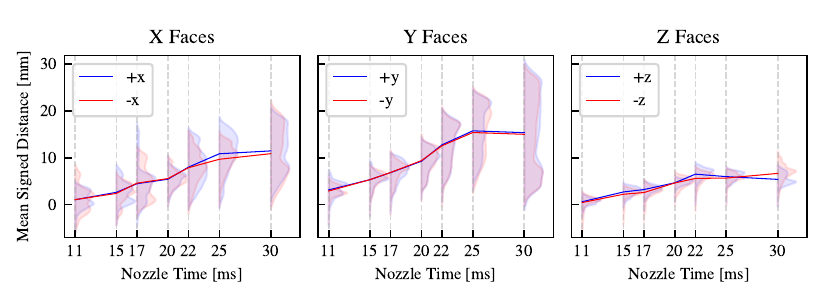}
\caption{Mean signed surface distance from CAD (solid lines) per face group ($\pm x$, $\pm y$, $\pm z$) over increasing nozzle times (\qtyrange[range-units = single, range-phrase = --]{11}{30}{\milli\s}). Additionally, the underlying distribution of point-wise distances is shown as a kernel density estimate (KDE) in the background for each nozzle time setting.}
\label{fig:results:deviations_by_face_group_vs_nozzle_time}
\end{figure}

To investigate the relationship between water dosage and geometric deviation, \Cref{fig:results:deviations_by_face_group_vs_nozzle_time} presents the mean signed surface distance for each face group ($x$, $y$, $z$) across a range of nozzle times (\qtyrange[range-units = single]{11}{30}{\milli\s}). Each subplot shows the average deviation separately for the two opposing faces in each direction (e.g., $+x$ and $-x$), with shaded regions representing the underlying distribution of point-wise distances as a kernel density estimate (KDE).

\begin{figure}[ht!]
\centering
\includegraphics[width=0.4\textwidth,clip,trim = 0 4cm 0 4cm]{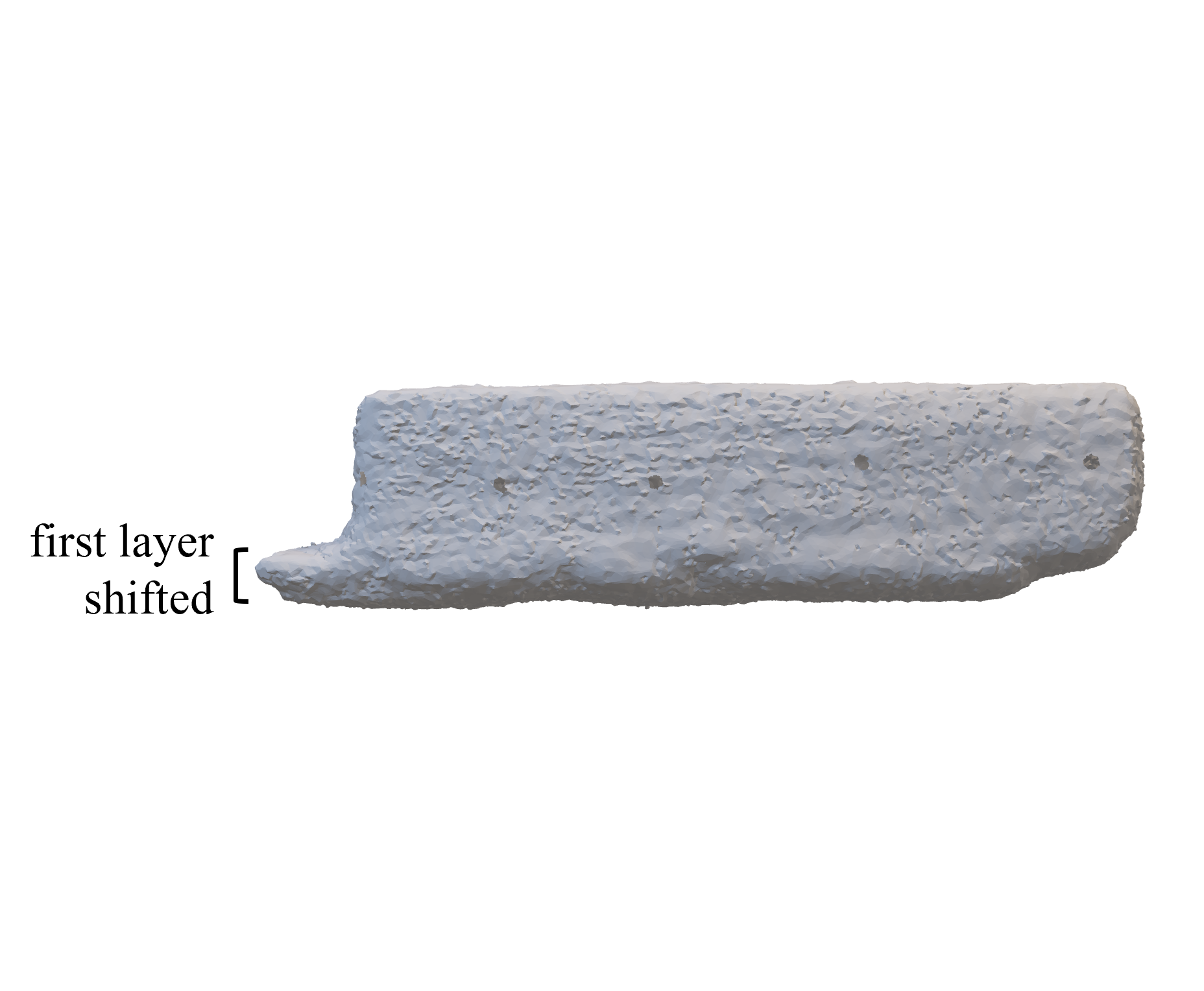}
\caption{Side view of a prism displaying typical deformation of the first layer due to low friction on the build plate during recoating. This specific sample is an extreme case and was not included in the main analysis.}
\label{fig:3_STL_nose}
\end{figure}

A clear trend can be observed for the $x$ and $y$ faces, where the mean signed distance increases strongly with higher water dosage. This indicates that the printed prisms expand laterally as the water content rises. The $z$ faces show a similar but less pronounced growth, suggesting that the vertical deviation is smaller compared to the horizontal directions. There is generally no significant difference between the positive and negative directions of each axis, both in terms of the mean value and the overall distribution. Minor differences in the KDEs arise from local surface features, such as irregularities in the first printed layer that cause deviations of a few millimeters. The main issue remains the low friction of wetted powder on the base plate of the printer, which allows the printer to drag the first layer along its movement direction when recoating (see \Cref{fig:3_STL_nose}).

At higher water contents, a stronger asymmetry between the $+z$ and $-z$ faces becomes visible. Despite their near-symmetric geometry and similarly restricted diffusion conditions in both vertical directions (the base plate limits movement in the $-z$ direction, while limited powder redeposition restricts movement in the $+z$ direction), gravity appears to influence water droplet behavior at higher dosages, causing preferential downward migration that leads to higher deviations on the bottom face. The KDEs across all faces are not fully symmetric, which can be attributed to localized outliers in both directions. These outliers are mainly caused by surface indentations, layer irregularities or handling errors during excavation.

The KDEs also widen in $x$ and $y$ direction (greater variance) with increasing water dosage, indicating greater variability and reduced process stability at higher nozzle times. This highlights the need for tighter process control or compensation strategies in applications requiring high dimensional accuracy.

\begin{figure}[ht!]
\centering
\includegraphics[clip, trim=0.28cm 0.29cm 0.28cm 0.29cm]{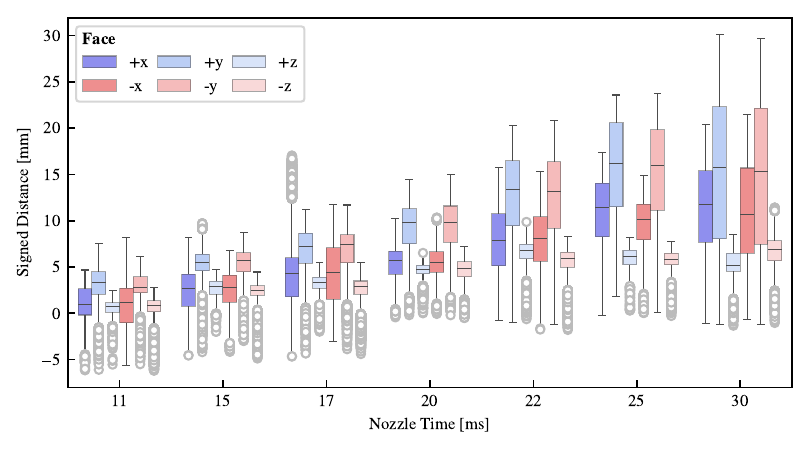}
\caption{Boxplot of mean signed surface distance from CAD per face group ($\pm x$, $\pm y$, $\pm z$) over increasing water dosages, including median and outliers (lower outliers: below $<Q_1-1.5\cdot \mathrm{IQR}$, upper outliers: $>Q_3+1.5\cdot \mathrm{IQR}$, with $Q_1$, $Q_3$ and $\mathrm{IQR}$ being the first and third quartiles and inter quartile range, respectively).}
\label{fig:3_boxplot}
\end{figure}

Another visualization of the relationship between water dosage and geometric deviation is provided in \Cref{fig:3_boxplot}. It is clearly visible that the number of outliers is numerous, especially in the lower water dosage range. This is mainly due to surface irregularities and layer defects that occur more frequently at lower water contents, where the powder bed is less stable. At higher water contents ($\geq \SI{20}{\milli\s}$), the frequency of outliers diminishes and concentrates below the mean, indicating that the overall integrity of the printed parts improved and defects during excavation or handling decreased. Also there is a notable jump in minimum deviations going from \SI{17}{\milli\s} to \SI{20}{\milli\s}, which suggests a threshold effect where excessive water leads to significant geometric distortion.

\begin{figure}[ht!]
	\captionsetup{justification=centering}
    \centering
    \begin{subfigure}[b]{0.32\textwidth}
        \includegraphics[clip, trim=0.3cm 0.3cm 0.3cm 0.3cm,width=1\textwidth]{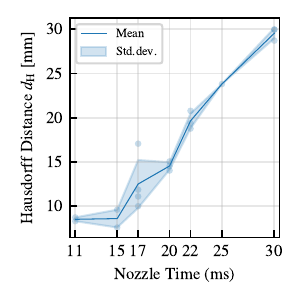}
        \caption{}
        \label{fig:3_metrics_hausdorff}
    \end{subfigure}
	\hfill
    \begin{subfigure}[b]{0.32\textwidth}
        \includegraphics[clip, trim=0.3cm 0.3cm 0.3cm 0.3cm,width=1\textwidth]{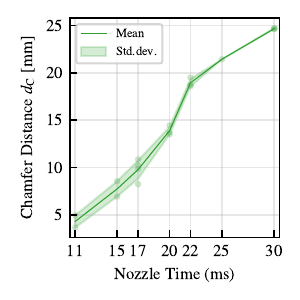}
        \caption{}
        \label{fig:3_metrics_chamfer}
    \end{subfigure}
	\hfill
    \begin{subfigure}[b]{0.32\textwidth}
        \includegraphics[clip, trim=0.3cm 0.3cm 0.3cm 0.3cm,width=1\textwidth]{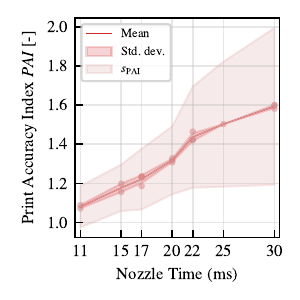}
        \caption{}
        \label{fig:3_metrics_pai}
    \end{subfigure}
    \caption{Evaluated metrics over nozzle time. a) Hausdorff distance b) Chamfer distance c) Print accuracy index (PAI) with the standard deviations of all samples in each nozzle time group and of all values across all samples per nozzle time group (according to \Cref{eqn:2_s_PAI}).}
    \label{fig:3_metrics}
\end{figure}

For a broader assessment of the geometric deviations, we additionally calculated the scalar metrics that were introduced in \Cref{sec:2_metrics}. The results are presented in \Cref{fig:3_metrics} where the mean values of all three metrics are plotted together with the corresponding standard deviations for increasing nozzle times. All metrics show a positive correlation with the water content, increasing progressively as the water dosage rises. The Hausdorff distance (\Cref{fig:3_metrics_hausdorff}) exhibits a distinctly large standard deviation for the nozzle time of \SI{17}{\milli\s} because it is sensitive to the most extreme local deviation and this group contains samples with strong variations in the first printed layer. This behaviour is not visible in the other two metrics because they are based on averaging procedures. The Chamfer distance (\Cref{fig:3_metrics_chamfer}) and the print accuracy index (\Cref{fig:3_metrics_pai}) show comparable qualitative patterns for this reason. For all nozzle times, the values for the print accuracy index are larger than one which indicates that the printed objects are consistently larger than the reference prism. With regard to geometric acceptability, we observe Hausdorff distances of about \SI{8}{\milli\m} and Chamfer distances of about \SI{4}{\milli\m} for the samples with the lowest water content. Even these values imply that the shape accuracy is already limited even for very small water amounts and the situation worsens for higher water dosages. The standard deviation for the print accuracy index (over all samples in a nozzle opening time group) introduced in \Cref{eqn:2_s_PAI} is relatively large across all nozzle opening times, but increases with higher water dosage, indicating greater variability in dimensional accuracy at elevated water contents.

\subsection{Spatial Distribution of Signed Distances on Prism Faces}

\begin{figure}[ht!]
\centering
\includegraphics[clip,trim=0.0cm 0.0cm 0.0cm 0.0cm]{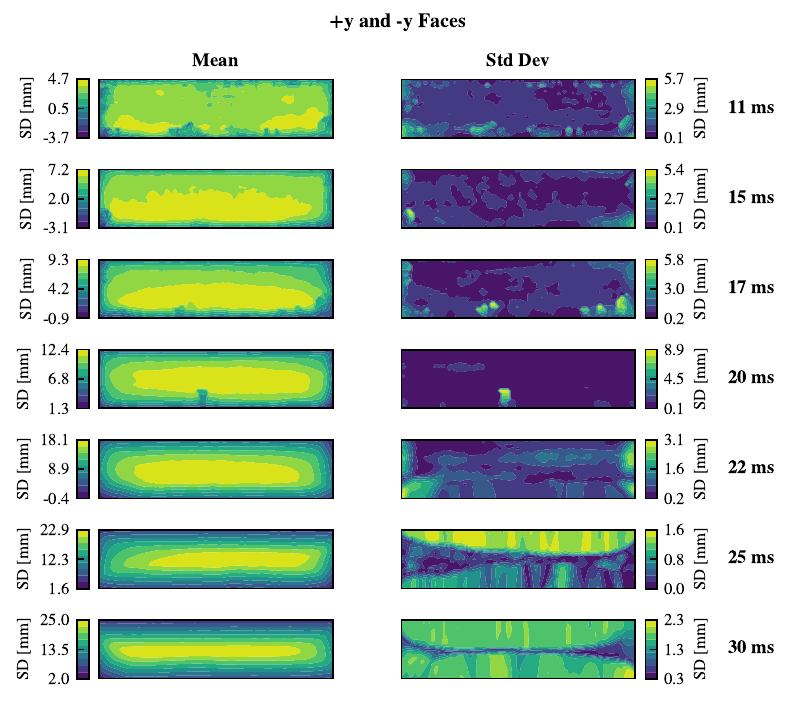}
\caption{Spatial distribution of mean signed distances and standard deviations for $y$-faces over different nozzle times. Each row corresponds to one nozzle time, with increasing water dosage from top to bottom.}
\label{fig:3_contour_y}
\end{figure}

To complement the global deviation metrics, the spatial distribution of the signed distances was evaluated for all samples and for all three face sets of the printed prisms. \Cref{fig:3_contour_y} presents the mean signed distance fields and the corresponding standard deviations for the faces aligned with the positive and negative $y$ directions. The data for the other directions is found in \ref{sec:A_Appendix}, see \Cref{fig:3_contour_x} for the $x$ and \Cref{fig:3_contour_z} for the $z$ direction. Each row corresponds to one nozzle time and therefore to one level of water dosage. For each nozzle time, the signed distance maps from all available samples were aggregated (in accordance with \Cref{sec:2_projection}).

The mean maps reveal a clear trend of increasing positive geometric deviation with increasing water dosage for $x$ and $y$ faces. At low nozzle opening times, deviations remain small and spatially unstructured, whereas higher nozzle opening times generate distinct circular/ellipsoidal deviation zones across broad regions of the same faces. The effect is consistent with the global metrics where a larger water content produced stronger swelling of the printed geometry. Especially for the $x$ and $y$ faces, we observe the overall highest mean values consistently in the center of the face, except for the lowest nozzle times. This suggests that excess water within densely surrounded voxel regions accumulates and promotes outward expansion. In contrast, voxels with fewer neighbours or reduced confinement exhibit smaller deviations, which supports the assumption that local drainage paths and material confinement govern the spatial distribution of shape expansion. The shapes become increasingly rounded for longer nozzle openings which supports the earlier observation that the prisms expand beyond the nominal reference volume. In $z$ direction, the effect is less pronounced but still visible. Here, the mean signed distances remain relatively small and similarly distributed even for high water dosages, indicating that vertical deformation is limited compared to horizontal spreading (as already mentioned in \Cref{sec:3_water_dosage_vs_geometry}).

The standard deviation maps in the right column of the contour plots reflect both systematic effects of the printing process and sample specific defects that become particularly visible when only few samples are available for averaging. For the $x$ faces, a recurring horizontal band of increased deviations appears in the lower third of the plots, which results from a shifted first layer present in some prisms and partially propagated into the adjacent layer (see \Cref{fig:3_STL_nose}). Quantitatively, the $x$ face standard deviations show no clear increase with higher nozzle times, which suggests that the printer exhibits a consistent layerwise deposition behaviour across the investigated range of water contents. The $y$ faces exhibit isolated large standard deviation values for nozzle times of \SI{17}{\milli\second} and \SI{20}{\milli\second}, caused by local nearest point mismatches during the signed distance computation (see discussion in \Cref{sec:2_projection}); for lower nozzle opening times, the deviations are more evenly distributed with local extrema near the edges, which is plausible given that edges are most susceptible to excavation related damage. Across all configurations with higher water contents, the $y$ faces show a characteristic separation between upper and lower regions with reduced deviations in the centre, consistent with the fact that the largest mean geometric deviations are located in that central zone. For the $z$ faces, elevated standard deviations again cluster near the boundaries, especially towards the $x$ faces, which can also be attributed to shifted first layers affecting the entire side surface. A visible diagonal band in the \SI{20}{\milli\second} group originates from one sample with a noticeable indentation in that region. While the limited sample size constrains the statistical power of the standard deviation analysis, the spatial patterns observed provide valuable preliminary insights into surface quality variations.

\subsection{Correlation between Water Content and Mechanical Response}

\begin{figure}[ht!]
	\captionsetup{justification=centering}
    \centering
    \begin{subfigure}[b]{0.445\textwidth}
        \includegraphics[width=1\textwidth]{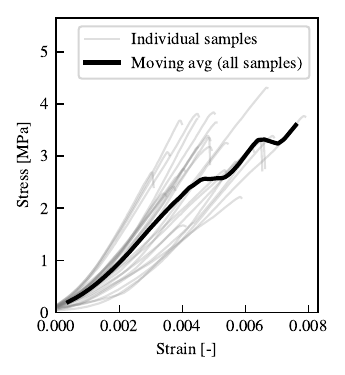}
        \caption{}
        \label{fig:3_mean_and_std_by_water_amount}
    \end{subfigure}
	\hfill
    \begin{subfigure}[b]{0.54\textwidth}
        \includegraphics[width=1\textwidth]{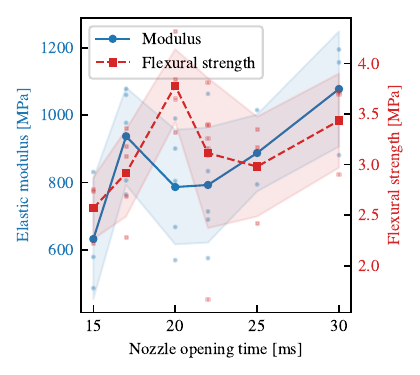}
        \caption{}
        \label{fig:3_mod_and_strength_vs_water_amount}
    \end{subfigure}
    \caption{Material tests a) evolution of the stress-strain curves of all samples including a moving average, b) flexural strength and Young's modulus vs. water dosage with mean values (solid line) and standard deviation (filled area).}
    \label{fig:3_mechanical_results}
\end{figure}

The determination of flexural strength and Young's modulus was carried out as described in \Cref{sec:2_mechanical_testing}. \Cref{fig:3_mean_and_std_by_water_amount} presents the stress-strain relationships for all tested samples and all nozzle times. Additionally, a moving average was computed for each nozzle time. All tested samples show a comparable overall behavior, with similar curve shapes and characteristic regions. There are some slight differences for the slopes of the curves (proportional to Young's modulus) and larger differences in the maximum stress values (approx. $\SI{2.5}{\mega\pascal}$ difference between the lowest and highest strength).

\Cref{fig:3_mod_and_strength_vs_water_amount} summarizes the flexural strengths and Young's moduli computed from the stress-strain curves (see \Cref{sec:2_mechanical_testing}) as a function of nozzle opening time, including their standard deviations. The data show relatively large variations within each group, particularly for strength, and no distinct optimum water content can be identified when considering both stiffness and flexural strength. This indicates that water dosage does not fundamentally alter the mechanical response of the printed material within the investigated range. The samples printed with a nozzle time of \SI{15}{\milli\second} exhibit the lowest average stiffness and strength, while those printed with \SI{17}{\milli\second} and \SI{30}{\milli\second} show higher values despite their very different water contents but similar water-to-cement ratios. The highest mean flexural strength was obtained for the \SI{20}{\milli\second} samples, which might be attributed to a favorable water-to-cement ratio that promotes sufficient hydration while avoiding excessive porosity. Excluding the \SI{20}{\milli\second} samples from the analysis would result in a slightly increasing trend in strength with rising water content, and likewise, removing the \SI{17}{\milli\second} data would reveal a similar trend for stiffness. These observations suggest that both groups may represent statistical outliers, which is plausible given the small sample size and the considerable data spread.

From a material point of view, a moderate increase in both stiffness and strength with increasing water content would be expected. A higher water dosage improves the local bonding of the powder particles, reducing the occurrence of unbound or dry regions that do not contribute to load-bearing capacity. However, as demonstrated in the previous sections, the water-to-cement ratio remains relatively constant across the tested nozzle times. This suggests that additional water primarily spreads over a larger volume rather than significantly altering the local material composition. Consequently, the mechanical properties remain largely unaffected by the investigated range of water dosages. Excessive water would still be expected to promote pore formation upon evaporation, which could reduce both stiffness and strength, though this effect was not observed within the tested parameter range.

\begin{figure}[ht!]
\centering
\includegraphics[clip,trim=0.0cm 0.0cm 0.0cm 0.0cm]{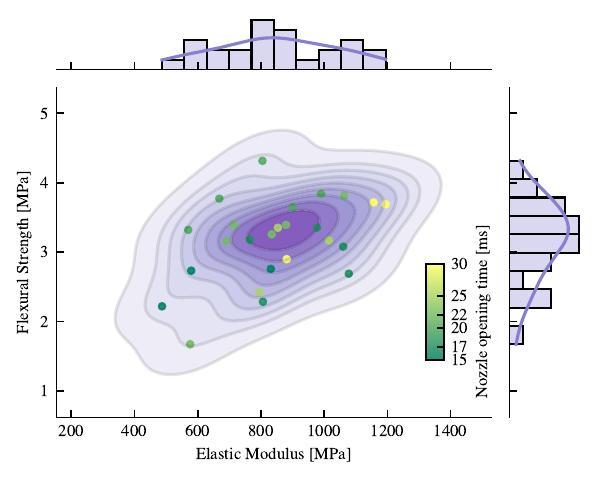}
\caption{Joint KDE of flexural strength vs. Young's modulus including marginal distributions with individual samples color coded by nozzle time (\mycirc[nozzle_green] green = low, \mycirc[nozzle_yellow] yellow = high)}
\label{fig:3_joint_mod_strength}
\end{figure}

An evident correlation between stiffness and strength cannot be observed for the tested specimens, as higher stiffness does not necessarily coincide with higher strength (see \Cref{fig:3_joint_mod_strength}). Most results cluster around a stiffness-strength pair of approximately \SI{900}{\mega\pascal} and \SI{3.2}{\mega\pascal}. Some variation is observed across different nozzle opening times: samples printed with \SI{20}{\milli\second} and \SI{22}{\milli\second} tend toward higher flexural strength with average stiffness, while the \SI{17}{\milli\second} and \SI{30}{\milli\second} samples show higher stiffness but moderate strength. The weakest performance in terms of both parameters is found for the \SI{15}{\milli\second} samples, whereas the \SI{20}{\milli\second} and \SI{30}{\milli\second} nozzle opening times yield the highest combined results. However, given the considerable scatter in the data and the limited sample size, it remains unclear whether these differences represent genuine material trends or statistical variation.

\subsection{Compensation Study}

To explore potential strategies for improving geometric fidelity in powder bed 3D printing of cementitious materials, a first proof of concept was carried out to compensate known geometric uncertainties. The aim was to demonstrate that systematic modification of the voxel geometry can significantly enhance the dimensional accuracy of printed components.

In this preliminary trial, the compensation approach was applied to the same prism geometry with nominal dimensions of \qtyproduct[allow-quantity-breaks]{159.6 x 39.9 x 39.9}{\milli\m}, printed using the highest water dosage corresponding to a nozzle time of \SI{30}{\milli\second}. The geometric analysis of these specimens, as discussed in earlier sections, revealed pronounced positive deviations in all directions, indicating an overall oversizing of the printed prism. Based on the signed distance data from these samples, localized corrections were introduced directly in the voxel model. For regions exhibiting the largest positive deviations, two voxels were removed along the direction of the deviation, while regions with moderate deviations were corrected by removing a single voxel.

\begin{figure}[ht!]
\centering
\includegraphics[clip,trim=0.7cm 0.4cm 0.3cm 0.18cm]{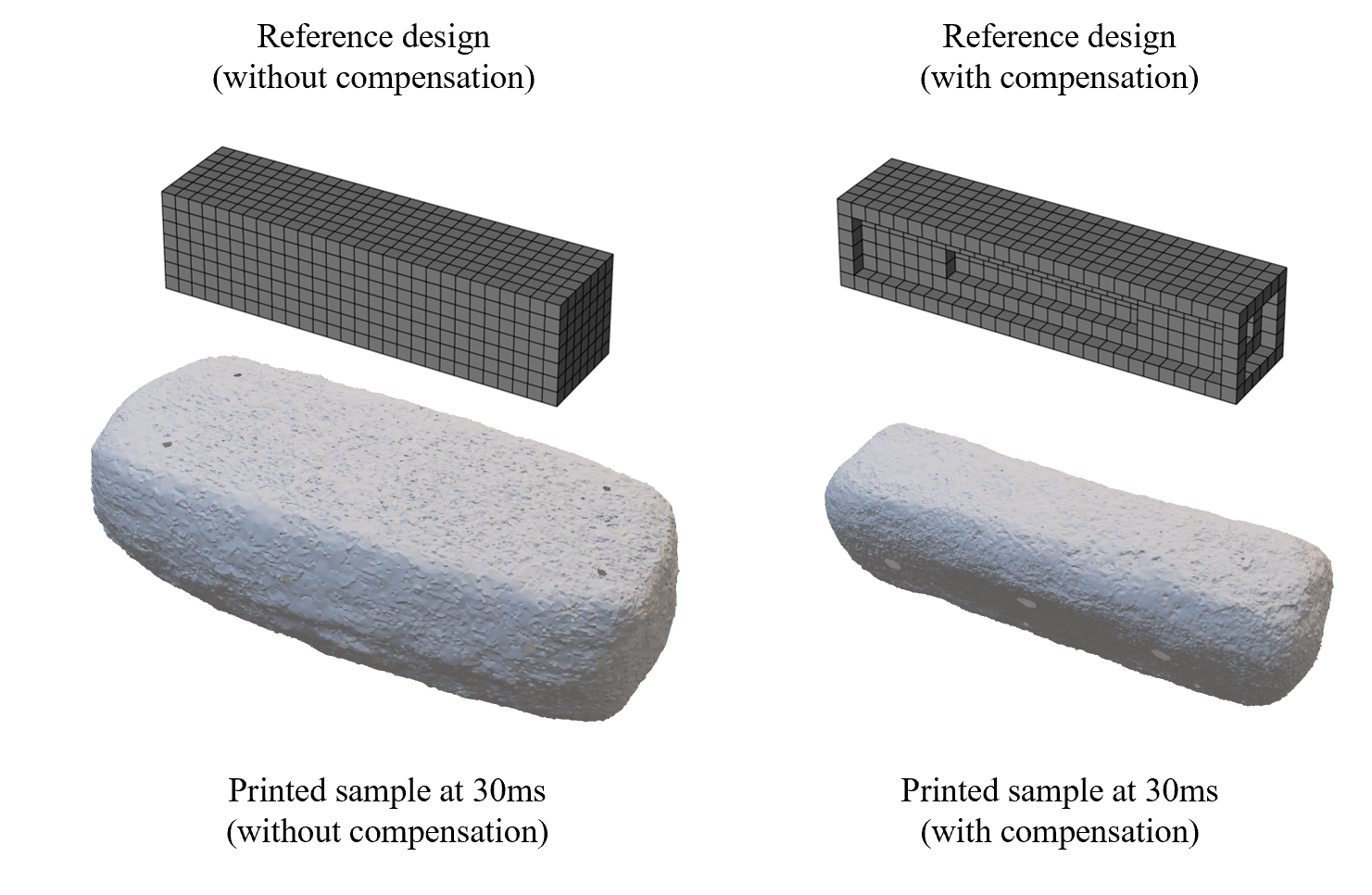}
\caption{Adapting the shape of the prism based on the signed distance data from the initial \SI{30}{\milli\second} print. The reference design (top left) was modified by removing voxels in regions with high positive deviations, resulting in a compensated design (top right). The printed samples for both cases are shown in the bottom row.}
\label{fig:3_compensation}
\end{figure}

In addition to these local corrections, a global adjustment was introduced. Because the original prisms exhibited systematic oversizing even along the build direction, one layer of voxels was removed from each dimension. The base geometry thus changed from a grid of 28 x 7 x 7 voxels voxels to 27 x 6 x 6 voxels. The resulting voxel configuration after all local and global adaptations is illustrated in \Cref{fig:3_compensation} (top right).

\begin{figure}[ht!]
\centering
\includegraphics[clip,trim=1.0cm 1.8cm 0.5cm 1.3cm]{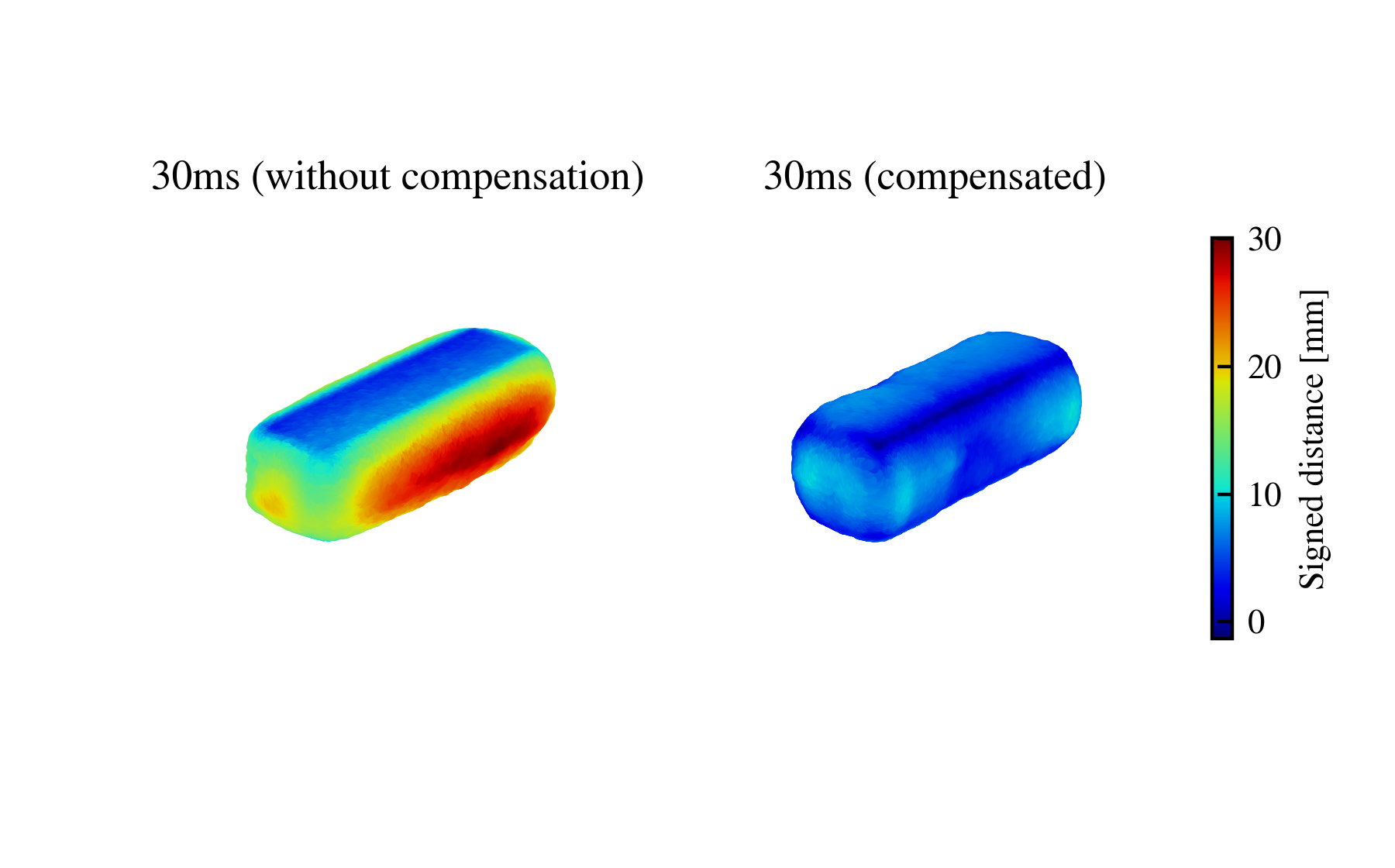}
\caption{Comparison of signed distance distributions between the initial \SI{30}{\milli\second} nozzle time print and the compensated design. The compensated specimen shows a significantly reduced maximum deviation and a more overall rectangular appearance, indicating improved geometric fidelity.}
\label{fig:3_SD comparison}
\end{figure}

The adapted prism was printed using identical process parameters. The resulting specimen displayed a significantly reduced overall size and improved dimensional conformity, yielding a geometry closer to the intended rectangular profile than the uncompensated version (see bottom right of \Cref{fig:3_compensation}). Quantitative evaluation of the new geometry, conducted through the same scanning and alignment procedure described earlier, confirmed a substantial improvement in geometric accuracy. The majority of signed distances were within the range of a single voxel size, with only isolated local deviations reaching up to \SI{10}{\milli\meter}, as visualized in \Cref{fig:3_SD comparison}.

This experiment demonstrates that targeted voxel-level modifications based on prior geometric deviation data can effectively compensate for systematic inaccuracies in powder bed 3D printing. Future work will build upon these findings to develop a predictive framework capable of automatically adjusting input CAD geometries based on process-specific deformation patterns (similar to \cite{Jadayel2025_ARTICLE}, but without a printing and scanning a sacrifical print first). The envisioned tool aims to estimate the final as-printed shape from the nominal design and to iteratively modify the geometry such that the manufactured parts exhibit minimal deviations from the intended dimensions.

\section{Conclusions and outlook}
\label{sec:4_conclusions_and_outlook}

This study systematically investigated the influence of water dosage on the geometric fidelity and mechanical properties of cementitious specimens produced by powder bed 3D printing. Prismatic specimens were fabricated using a printer with varying nozzle opening times ranging from \SI{11}{\milli\s} to \SI{30}{\milli\s}, corresponding to different water contents. The printed geometries were characterized through high-resolution 3D scanning, ICP-based alignment, and comprehensive deviation analysis, while mechanical performance was evaluated via three-point bending and compression tests.

The geometric analysis revealed a strong correlation between water dosage and dimensional accuracy. Specimens printed with higher water contents exhibited progressively increasing deviations from the nominal CAD geometry, with mean signed distances rising from approximately \SI{3}{\milli\m} at \SI{11}{\milli\s} to over \SI{15}{\milli\m} at \SI{30}{\milli\s} for the $x$ and $y$ faces. This expansion was primarily attributed to lateral water spreading through the loose powder bed, resulting in pronounced edge rounding and cross-sectional deformation. The $z$-faces showed comparatively smaller deviations, indicating that vertical constraints during layer deposition effectively limited upward and downward material displacement. Direction-dependent analysis confirmed that geometric distortion was most severe in the horizontal plane.

Three complementary methods were employed to estimate the water-to-cement ratio of the printed structures. While the theoretical voxel-based approach yielded unrealistically high values exceeding unity for high water contents, both the mass-based and geometric volume-corrected methods produced more plausible ratios in the range of \num{0.2} to \num{0.4}. Importantly, these ratios remained relatively constant across the tested nozzle times, suggesting that additional water primarily spread over a larger volume rather than significantly altering the local material composition. Therefore, the overall geometric deviations increase with water dosage. This finding also provides a partial explanation for the observed mechanical behavior.

Mechanical testing demonstrated that flexural strength and Young's modulus remained largely independent of water dosage within the investigated parameter range. Mean flexural strengths varied between \SI{2.5}{\mega\pascal} and \SI{3.5}{\mega\pascal}, while mean stiffness values ranged from approximately \SI{600}{\mega\pascal} to \SI{1100}{\mega\pascal}. No clear optimum water content could be identified, and considerable scatter was observed within each nozzle time group. The absence of a strong correlation between water dosage and mechanical properties aligns with the finding that the water-to-cement ratio remained relatively stable. Moderate increases in both strength and stiffness with rising water content, potentially due to improved particle bonding and reduced dry regions, were partially masked by statistical variation and the limited sample size.

The proof-of-concept compensation study successfully demonstrated that targeted voxel-level modifications based on prior geometric deviation data can substantially improve dimensional accuracy. By removing voxels from regions with the largest positive deviations and reducing the overall layer count, the adapted geometry exhibited significantly reduced signed distances, with the majority of deviations confined to within a single voxel size. This preliminary result validates the feasibility of compensation strategies and establishes a foundation for developing predictive tools capable of automatically correcting input geometries to account for process-specific deformation patterns.

In summary, this work provides quantitative evidence that water dosage in powder bed 3D printing of cementitious materials exerts a dominant influence on geometric fidelity while having only a minor effect on mechanical performance within the tested range. The systematic characterization of deviation patterns and the successful implementation of a compensation approach offer valuable insights for optimizing process parameters and advancing the dimensional accuracy of powder bed printed concrete components.

Building on the findings of this study, several avenues for future research and development can be identified to advance the field of powder bed 3D printing for cementitious materials.
First, develop an automated compensation method that predicts printed geometry from CAD data and process settings. This should evolve from the manual voxel edits shown here and use machine learning and physics based models to estimate deformation and adjust designs to enable accurate one shot printing.
Second, increase sample size to strengthen the statistical power of the mechanical testing results and enable more robust conclusions regarding the influence of water dosage on material properties.
Finally, apply the geometric analysis method to more complex shapes with varied features to study the influence of geometry on deviation patterns and compensation effectiveness.

\appendix

\section*{CRediT authorship contribution statement}

\textbf{Christoph Wolf:} Conceptualization; Data curation; Formal analysis;
Investigation; Methodology; Software; Validation; Visualization; Writing -- original draft; Writing -- review \& editing.
\textbf{Petr Hlaváček:} Investigation; Methodology; Resources; Writing -- review \& editing.
\textbf{Annika Robens-Radermacher:} Conceptualization, Methodology; Writing -- review \& editing.
\textbf{Daniel Kadoke:} Data curation; Investigation; Resources.
\textbf{Jörg F. Unger:} Conceptualization, Funding acquisition; Methodology; Project administration; Supervision; Writing -- review \& editing.

\section*{Declaration of Competing Interest}

The authors declare that they have no known competing financial interests or personal relationships that could have appeared to influence the work reported in this paper.

\section*{Acknowledgements}

The authors gratefully acknowledge the financial support by the German Federal Ministry for Economic Affairs and Energy (BMWK) within the collaborative project \glqq{}SupLeichtAF - Structure-Optimized Lightweight Constructions through Additive Manufacturing in the Powder Bed\grqq{} (Grant number 03LB3031A).

\section*{Declaration of generative AI and AI-assisted technologies in the manuscript preparation process}

During the preparation of this work, the authors used Microsoft Copilot (GPT-4o and other models as of 2025) and ChatGPT (GPT-5) in order to improve the clarity and correctness of English grammar and phrasing in the manuscript. After using this tool/service, the authors reviewed and edited the content as needed and take full responsibility for the content of the published article. Furthermore Github Copilot (with the models GPT-5, Claude Sonnet 4 and 4.5) was used to support the writing of code for algorithmic optimizations, data analysis and visualization. All scientific content, ideas and interpretations originate from the authors.

\clearpage
\section{Appendix}\label{sec:A_Appendix}

\begin{figure}[ht!]
\centering
\includegraphics[clip,trim=0.0cm 0.0cm 0.0cm 0.0cm]{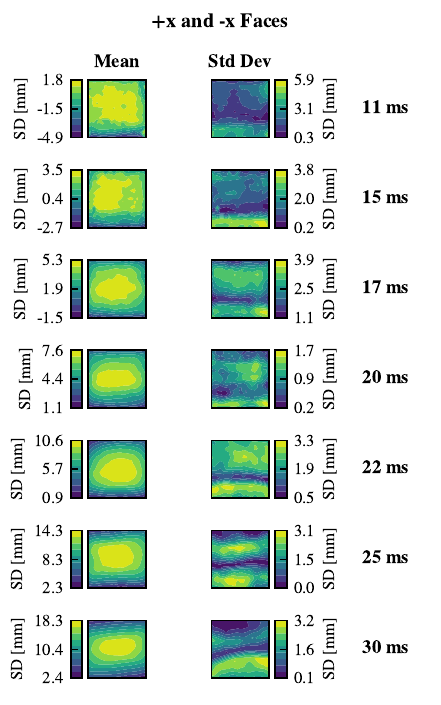}
\caption{Spatial distribution of mean signed distances and standard deviations for $x$-faces over different nozzle times. Each row corresponds to one nozzle time, with increasing water dosage from top to bottom.}
\label{fig:3_contour_x}
\end{figure}

\begin{figure}[ht!]
\centering
\includegraphics[clip,trim=0.0cm 0.0cm 0.0cm 0.0cm]{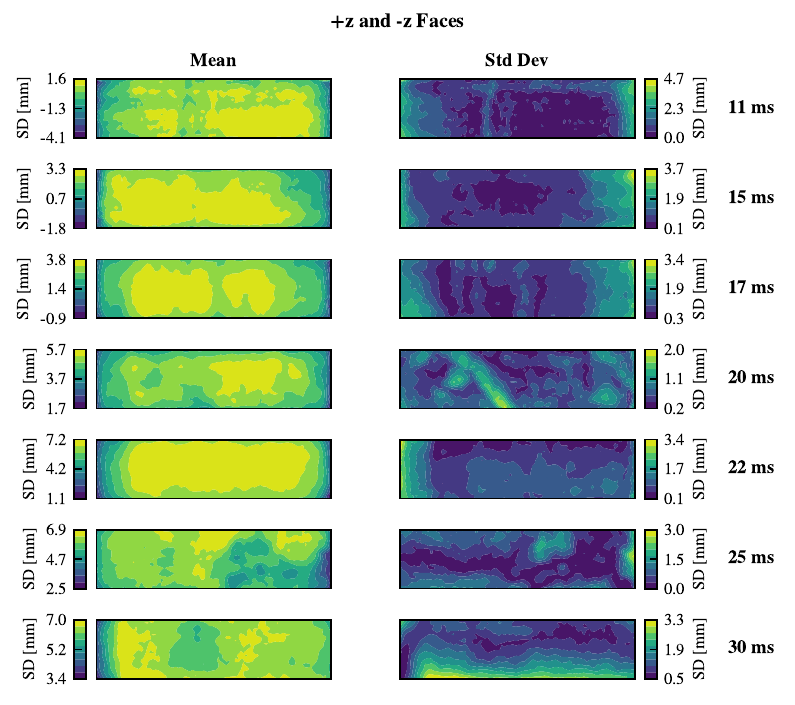}
\caption{Spatial distribution of mean signed distances and standard deviations for $z$-faces over different nozzle times. Each row corresponds to one nozzle time, with increasing water dosage from top to bottom.}
\label{fig:3_contour_z}
\end{figure}

\clearpage
\section*{Data availability}

Data will be made available on request.
\label{sec:sample:appendix}

\bibliographystyle{elsarticle-num} 
\bibliography{2024_11_powderbed_uncertainties}{}

\end{document}